\documentclass{aa} 
\usepackage{graphicx}
\begin{document}

%
\def\ffam {\hbox{$\,.\!\!^{\prime}$}}
\def\ffas {\hbox{$\,.\!\!^{\prime\prime}$}}
\def\ffs {\hbox{$\,.\!\!^{\rm s}$}}
\def\ffm {\hbox{$\,.\!\!^{\rm m}$}}
\def\HI {H\kern0.1em{\sc i}} 
\def\txs {TXS\,2226{\tt -}184}
\def\pks {PKS\,2322{\tt -}123}
\def\radm {rad m$^{-2}$} 
\def\ab {$\sim$}
\def\etal {{\sl et~al.\ }}
\def\eg {{\it e.~g.}}
\def\ie {{\it i.~e.}}
\def\dg{$^{\circ}$}
\def\kms{km s$^{-1}$}
\def\solmass {\hbox{M$_{\odot}$}}
\def\solum {\hbox{L$_{\odot}$}}

\def \la{\mathrel{\mathchoice   {\vcenter{\offinterlineskip\halign{\hfil
$\displaystyle##$\hfil\cr<\cr\sim\cr}}}
{\vcenter{\offinterlineskip\halign{\hfil$\textstyle##$\hfil\cr
<\cr\sim\cr}}}
{\vcenter{\offinterlineskip\halign{\hfil$\scriptstyle##$\hfil\cr
<\cr\sim\cr}}}
{\vcenter{\offinterlineskip\halign{\hfil$\scriptscriptstyle##$\hfil\cr
<\cr\sim\cr}}}}}

\def \ga{\mathrel{\mathchoice   {\vcenter{\offinterlineskip\halign{\hfil
$\displaystyle##$\hfil\cr>\cr\sim\cr}}}
{\vcenter{\offinterlineskip\halign{\hfil$\textstyle##$\hfil\cr
>\cr\sim\cr}}}
{\vcenter{\offinterlineskip\halign{\hfil$\scriptstyle##$\hfil\cr
>\cr\sim\cr}}}
{\vcenter{\offinterlineskip\halign{\hfil$\scriptscriptstyle##$\hfil\cr
>\cr\sim\cr}}}}}

\newcommand\nodata{ ~$\cdots$~ }

\title{New H$_2$O masers in Seyfert and FIR bright galaxies
       \thanks{Based on observations with the 100-m telescope of the MPIfR 
       (Max-Planck-Institut f{\"u}r Radioastronomie) at Effelsberg.}}

\author{C. Henkel\inst{1} \and 
        A. B. Peck\inst{2} \and 
        A. Tarchi\inst{3,4} \and
        N. M. Nagar\inst{5,6,7} \and 
        J. A. Braatz\inst{8} \and 
        P. Castangia\inst{4,9} \and
        L. Moscadelli\inst{4}}

\offprints{C. Henkel, \email{chenkel@mpifr-bonn.mpg.de}}

\institute{Max-Planck-Institut f{\"u}r Radioastronomie, Auf dem 
           H{\"u}gel 69, D-53121 Bonn, Germany
           \and 
           Harvard-Smithsonian Center for Astrophysics, SAO/SMA Project,
           654 N. A'ohoku Pl., Hilo, HI 96720, USA 
           \and
           Istituto di Radioastronomia, CNR, Via Gobetti 101, I-40129-Bologna, 
           Italy 
           \and 
           INAF-Osservatorio Astronomico di Cagliari, Loc. Poggio dei Pini,
           Strada 54, I-09012 Capoterra (CA), Italy
           \and 
           INAF, Arcetri Observatory, Largo E. Fermi 5, I-50125 Florence,
           Italy 
           \and 
           Kapteyn Instituut, Postbus 800, NL-9700 AV Groningen, The 
           Netherlands
           \and 
           Astronomy Group, Departamento de F\'{\i}sica,
           Universidad de Concepci{\'o}n, Casilla 160-C, Concepci{\'o}n, Chile 
           \and
           National Radio Astronomy Observatory, P.O. Box 2, Green Bank, 
           WV 24944, USA 
           \and 
           Universit{\'a} di Cagliari, Dipartimento di Fisica, Cittadella
           Universitaria, I-09012 Capoterra (CA), Italy}

\date{Received 14 October 2004 / Accepted 16 February 2005}


\abstract{Using the Effelsberg 100-m telescope, detections of four extragalactic water vapor masers are reported. 
 Isotropic luminosities are $\sim$50, 1000, 1, and 230\,L$_{\odot}$ for Mrk\,1066 (UGC~2456), Mrk~34, NGC~3556, 
 and Arp~299, respectively. Mrk~34 contains by far the most distant and one of the most luminous water vapor 
 megamasers so far reported in a Seyfert galaxy. The interacting system Arp~299 appears to show two maser hotspots 
 separated by approximately 20\arcsec. With these new results and even more recent data from Braatz et al. (2004), 
 the detection rate in our sample of Seyferts with known jet-Narrow Line Region interactions becomes 50\% (7/14), 
 while in star forming galaxies with high ($S_{\rm 100\mu m}$$>$50\,Jy) far infrared fluxes the detection 
 rate is 22\% (10/45). The jet-NLR interaction sample may not only contain `jet-masers' but also a significant 
 number of accretion `disk-masers' like those seen in NGC\,4258. A statistical analysis of 53 extragalactic H$_2$O 
 sources (excluding the Galaxy and the Magellanic Clouds) indicates (1) that the correlation between IRAS Point 
 Source and H$_2$O luminosities, established for individual star forming regions in the galactic disk, also holds 
 for AGN dominated megamaser galaxies, (2) that maser luminosities are not correlated with 60$\mu$m/100$\mu$m 
 color temperatures and (3) that only a small fraction of the luminous megamasers ($L_{\rm H_2O}$ $>$ 100\,L$_{\odot}$) 
 detectable with 100-m sized telescopes have so far been identified. The H$_2$O luminosity function (LF) suggests 
 that the number of galaxies with 1\,L$_{\odot}$ $<$ $L_{\rm H_2O}$ $<$ 10\,L$_{\odot}$, the transition range between 
 `kilomasers' (mostly star formation) and `megamasers' (active galactic nuclei), is small. The overall slope of the 
 LF, $\sim$--1.5, indicates that the number of detectable masers is almost independent of their luminosity. If the 
 LF is not steepening at very high maser luminosities and if it is possible to find suitable candidate sources, H$_2$O 
 megamasers at significant redshifts should be detectable even with present day state-of-the-art facilities. 
 \keywords{Masers -- Galaxies: active -- Galaxies: Jets -- Galaxies: Seyfert -- Galaxies: starburst -- Radio 
 lines: galaxies}}

\titlerunning{New H$_2$O masers in Seyferts and starbursts}
\authorrunning{C. Henkel et al.}

\maketitle

\section{Introduction and sample selection}

Extragalactic water vapor masers, observed through the 22\,GHz ($\lambda$$\sim$1.3\,cm) $J_{\rm K_aK_c}$ = 
6$_{16}-5_{23}$ line of ortho-H$_2$O that traces warm ($T_{\rm kin}$$\ga$400\,K) and dense ($n$(H$_2$)$\ga$10$^7$\,cm$^{-3}$)
molecular gas (e.g. Kylafis \& Norman 1987, 1991; Fiebig \& G{\"u}sten 1989), are primarily seen as a means to 
probe nuclear accretion disks in active galaxies. The best known source, NGC~4258, shows a thin, slightly warped, 
nearly edge-on Keplerian disk of subparsec scale enclosing a central mass of $\sim$4$\times$10$^{7}$\,M$_{\odot}$ (e.g. 
Greenhill et al. 1995; Miyoshi et al. 1995; Herrnstein et al. 1999). 

There is evidence, however, for additional classes of extragalactic H$_2$O masers. There are sources in which at 
least a part of the H$_2$O emission appears to be the result of an interaction between the nuclear radio jet and an 
encroaching molecular cloud (e.g. Mrk\,348; Peck et al. 2003). 

Most of the nuclear water vapor sources are characterised by (isotropic) $L_{\rm H_2O}$ $>$ 10\,L$_{\odot}$ and are 
classified as `megamasers'. H$_2$O masers associated with prominent star forming regions similar to those seen in the 
Galaxy (e.g. in M~33; Greenhill et al. 1993) are less luminous and comprise the majority of known `kilomasers' 
($L_{\rm H_2O}$ $\la$ 10\,L$_{\odot}$). 

Providing bright, almost point-like hotspots, H$_2$O masers are ideal probes for Very Long Baseline Interferometry (VLBI). 
A broad variety of astrophysical studies is possible. This includes the determination of geometric distances and 3-dimensional 
velocity vectors of galaxies, masses of nuclear engines, maps of accretion disks and physics of nuclear jet-molecular cloud 
interaction (for recent reviews, see Greenhill 2002, 2004; Maloney 2002; Henkel \& Braatz 2003; Morganti et al. 2004; Henkel et al. 
2005). 

So far, almost 1000 active galaxies have been surveyed. In order to detect a large number of strong maser sources that 
could help to elucidate the nuclear environment of their parent galaxies and their geometric distance, typical detection 
limits were several 10\,mJy or more, resulting in detection rates between zero (e.g. Henkel et al. 1998) and a few percent 
(e.g. Henkel et al. 1984; Braatz et al. 1996, Greenhill et al. 2002). The low detection rates are probably the result of 
the limited sensitivity of the surveys, rather than an intrinsic lack of extragalactic H$_2$O masers. The technology exists 
to do deeper searches; what is required is a set of criteria to narrow the list of candidates from all nearby galaxies to 
a manageable few. Here we present the results of two quite different, but equally successful, deep searches {(to estimate 
distances, $H_0$ = 75\,km\,s$^{-1}$\,Mpc$^{-1}$ is used, whenever possible, throughout the paper)}.

\paragraph{Sample 1:} 
The targets of the first sample (hereafter `jet-maser' sample) have been selected from a collection of Seyfert 
galaxies with declination $>-$30{\degr} in which both the inclination of the host galaxy and the linear extent of the 
radio source are known (Nagar \& Wilson 1999). To maximize the detection rate, we have chosen galaxies which either show 
evidence at optical wavelengths of interaction between the radio jet and clouds in the narrow line region, or have a
face-on ($i$$<$35\degr) galaxy disk and extended pc to 100-pc scale radio structures, possibly suggesting that both the 
disk of the galaxy and the radio jet should be fairly close to the plane of the sky. This combination of geometries 
increases the probability that the radio jet lies close to the disk of the galaxy, thus increasing the likelihood of 
interaction between the radio jet and the galaxy interstellar medium (ISM). Indeed, two of the four `jet-maser' sources 
known prior to this study, NGC~1068 (Gallimore et al. 2001) and Mrk~348 (Peck et al. 2003) appear in the parent sample 
and fit the above criteria. The complete list of 14 jet-maser target sources is given in Table~1.

\paragraph{Sample 2:} The second sample (hereafter `far infrared maser' or `FIR-maser' sample) is comprised of all galaxies 
with declination $>$--30{\degr} and IRAS point source flux density $S_{100\mu \rm m}$$>$50\,Jy (e.g. Fullmer \& Lonsdale
1989). The same criteria were already used by Henkel et al. (1986; hereafter HWB) to perform a first FIR-flux-based survey 
in which they detected a new maser in the galaxy IC~10. Recently we have detected masers in IC~342 (Tarchi et al. 2002a) 
and NGC~2146 (Tarchi et al. 2002b), which were previously undetected although they were included in the sample of HWB. We 
attribute this result to an occcasional flare (IC~342) and to an improvement in the Effelsberg receiver and backend systems 
(NGC~2146). This motivated us to re-observe the list of previously undetected targets. The list of sources was compiled 
using the IRAS Point Source Catalog (e.g. Fullmer \& Lonsdale 1989) and is shown in Table~2. There is a total of 45 sources,
the great majority of them being spiral galaxies. Two of these already known to show maser emission, IC~342 and NGC~2146, 
were reobserved.

\begin{table*}
\label{jets}
\begin{scriptsize}
\begin{center}
\caption{`Jet-maser' observations}
\begin{tabular}{llcrrrrccc}
\hline
Source$^{\rm a)}$ & Seyfert & R.A.    & \multicolumn{1}{c}{Dec.}    & \multicolumn{1}{c}{$V_{\rm sys}^{\rm b)}$} &
         \multicolumn{1}{c}{$V$--range} &  rms  & Channel & Epoch$^{\rm c)}$ \\
                  &                   &                             &                              &
                                        &       &   &  Width                 \\
       & (Type)  & (J2000) & \multicolumn{1}{c}{(J2000)} & \multicolumn{1}{c}{(c$z$)}              &
         (km\,s$^{-1}$)                 & (mJy) & (km\,s$^{-1}$) &       \\
\hline
\\
{\it Mrk\,348 (NGC\,262)}\,$^{\rm d)}$  &2  &00 48 47.1  &  31 57 25     & 4500 &             &    &     &      \\
{\it Mrk\,1157 (NGC\,591)}\,$^{\rm d)}$ &2  &01 33 31.2  &  35 39 56     & 4550 &  4050,5080  &  8 & 1.1 & 0302 \\
Mrk\,573                                &2  &01 43 57.8  &  02 20 59     & 5175 &  4700,5700  & 13 & 1.1 & 0302 \\
{\it NGC\,1068}\,$^{\rm d)}$            &1.8&02 42 40.7  &--00 00 48     & 1135 &             &    &     &      \\ 
{\bf Mrk\,1066}\,$^{\rm d)}$            &2  &02 59 58.6  &  36 49 14     & 3605 & see Table 3 &    &     &      \\
MCG\,8-11-11                            &1.5&05 54 53.6  &  46 26 21     & 6140 &  5650,6680  &  6 & 1.1 & 0302 \\
{\it Mrk\,3}\,$^{\rm d)}$               &2  &06 15 36.3  &  71 02 15     & 4050 &  3600,4480  &  5 & 4.2 & 0701 \\
                                        &   &            &               &      &  3570,4580  &  9 & 1.1 & 0302 \\
Mrk\,79                                 &1.2&07 42 32.8  &  49 48 35     & 6650 &  6160,7190  &  8 & 1.1 & 0302 \\
{\bf Mrk\,34}\,$^{\rm d)}$              &2  &10 34 08.6  &  60 01 52     &15140 & see Table 3 &    &     &      \\
NGC\,3516                               &1.2&11 06 47.4  &  72 34 07     & 2650 &  2160,3180  &  9 & 1.1 & 0302 \\
{\it NGC\,4151}$^{\rm d)}$              &1.5&12 10 32.6  &  39 24 21     & 1000 &     0,2000  &  3 & 4.3 & 0300 \\ 
NCG\,5135                               &2  &13 25 43.8  &--29 50 02     & 4110 &  3630,4650  & 18 & 1.1 & 0302 \\
Mrk\,270 (NGC 5283)                     &2  &13 41 05.7  &  67 40 21     & 3120 &  2640,3650  &  9 & 1.1 & 0302 \\
Mrk\,1126 (NGC\,7450)                   &1.5&23 00 47.8  &--12 55 06     & 3190 &  2700,3730  &  9 & 1.1 & 0302 \\
\\
\hline
\end{tabular}
\end{center}
a) Source names in italics: previously detected sources; bold-faced: newly detected sources \\
b) $V_{\rm sys}$: Systemic velocity \\
c) Epoch: given are the month (first two digits) and the year (last two digits) \\ 
d) Mrk\,348: Falcke et al. (2000); Mrk\,1157: Braatz et al. (2004); NGC\,1068: Claussen et al. (1984); 
Mrk\,1066: this paper, for a more recent independent detection see Braatz et al. (2004); Mrk\,3: Braatz et al. 
(2004); Mrk\,34: this paper; NGC\,4151: Braatz et al. (2004) \\ 
\end{scriptsize}
\end{table*}

\begin{table*}
\label{FIR}
\begin{scriptsize}
\begin{center}
\caption{`FIR-maser' observations}
\begin{tabular}{lccrrrccc}
\hline
Source$^{\rm a)}$   & R.A.     & Dec.          & $V_{\rm sys}$ & $S_{100\mu \rm m}^{\rm b)}$ &  
         $V$--range & rms      & Channel       & Epoch$^{\rm c)}$                       \\
                    &          &               &          &                             &
                    &          & Width         &                                        \\
                    & (J2000)  & (J2000)       & (c$z$)   & (Jy)                        & 
      (km\,s$^{-1}$ &   (mJy)  & (km\,s$^{-1}$)&                                        \\
\hline
                             &             &                &       &     &           &    &     &      \\
{\it{IC\,10}}\,$^{\rm d)}$   & 00 20 27.0  & $\:\,$59 17 29 & --350 &  71 &           &    &     &      \\
{\it{NGC\,253}}\,$^{\rm d)}$ & 00 47 33.1  &--25 17 18      &   240 &1045 &           &    &     &      \\
NGC\,660                     & 01 43 01.6  & $\:\,$13 38 35 &   850 & 104 &  620,1100 &  7 & 1.1 & 0601 \\
                             &             &                &       &     &    0,1780 & 10 & 4.2 & 0302 \\
NGC\,891                     & 02 22 33.4  & $\:\,$42 20 57 &   525 & 148 &  320,900  & 20 & 1.1 & 0601 \\
NGC\,972                     & 02 34 12.9  & $\:\,$29 18 48 &  1540 &  65 &  700,2470 &  9 & 4.3 & 0302 \\
                             &             &                &       &     &  860,2080 & 20 & 1.1 & 0302 \\
NGC\,1055                    & 02 41 45.4  & $\:\,$00 26 35 &  1000 &  60 &  800,1250 & 17 & 1.1 & 1203 \\
                             &             &                &       &     &  450,1600 & 17 & 1.1 & 0404 \\
Maffei\,2                    & 02 41 55.1  & $\:\,$59 36 11 &  --35 & 227 & --250,210 & 12 & 1.1 & 0601 \\
                             &             &                &       &     & --250,200 & 15 & 1.1 & 0302 \\
{\it{NGC\,1068}}\,$^{\rm d)}$& 02 42 40.7  &--00 00 48      &  1135 & 240 &           &    &     &      \\
NGC\,1084                    & 02 46 00.0  &--07 34 37      &  1405 &  55 &  930,1940 & 20 & 1.1 & 0103 \\
{\it{IC\,342}}\,$^{\rm d)}$  & 03 46 48.6  & $\:\,$68 05 46 &    40 & 128 & --190,300 & 12 & 1.1 & 0601 \\
                             &             &                &       &     & --150,300 &  9 & 1.1 & 0302 \\ 
UGC\,02855                   & 03 48 22.6  & $\:\,$70 07 57 &  1200 &  79 &  980,1460 &  9 & 1.1 & 0601 \\
NGC\,1569                    & 04 30 46.8  & $\:\,$64 51 02 &   100 &  52 & --100,190 & 12 & 1.1 & 0601 \\
{\it{NGC\,2146}}\,$^{\rm d)}$& 06 18 39.6  & $\:\,$78 21 19 &   900 & 187 &  500,1190 &  7 & 1.1 & 0302 \\
NGC\,2403                    & 07 37 37.6  & $\:\,$71 19 32 &   130 &  56 & --100,400 &  9 & 1.1 & 0601 \\
NGC\,2559                    & 08 17 06.0  &--27 27 27      &  1560 &  66 & 1080,2090 & 20 & 1.1 & 0103 \\
NGC\,2903                    & 09 32 10.1  & $\:\,$21 30 04 &   550 & 104 &  100,1100 &  7 & 4.2 & 0701 \\
{\it{NGC\,3034}}\,$^{\rm d)}$& 09 55 52.2  & $\:\,$69 40 47 &   200 &1145 &           &    &     &      \\
{\it{NGC\,3079}}\,$^{\rm d)}$& 10 01 57.8  & $\:\,$55 40 47 &  1120 &  89 &           &    &     &      \\
NGC\,3521                    & 11 05 48.6  &--00 02 09      &   850 &  85 &  320,1340 & 29 & 1.1 & 0302 \\
{\bf{NGC\,3556}}\,$^{\rm d)}$& 11 11 31.2  & $\:\,$55 40 25 &   700 &  61 &see Table 3&    &     &      \\
NGC\,3627                    & 11 20 15.0  & $\:\,$12 59 30 &   725 & 106 &  300,1250 &  8 & 4.2 & 0701 \\
NGC\,3628                    & 11 20 17.0  & $\:\,$13 35 20 &   842 & 103 &  400,1380 &  7 & 4.2 & 0701 \\
{\bf{Arp\,299}}\,$^{\rm d)}$ & 11 28 31.9  & $\:\,$58 33 45 &  3120 & 111 &see Table 3&    &     &      \\
NGC\,4038                    & 12 01 52.8  &--18 51 54      &  1640 &  76 & 1020,2300 & 28 & 1.1 & 0502 \\
NGC\,4088                    & 12 05 34.2  & $\:\,$50 32 21 &   750 &  52 &  280,1280 & 16 & 1.1 & 0302 \\
NGC\,4102                    & 12 06 23.1  & $\:\,$52 42 39 &   850 &  69 &  600,1100 & 17 & 1.1 & 0601 \\
NGC\,4254                    & 12 18 49.5  & $\:\,$14 25 00 &  2400 &  72 & 2000,2930 &  6 & 4.3 & 0701 \\
NGC\,4303                    & 12 21 54.9  & $\:\,$04 28 25 &  1565 &  62 & 1080,2100 & 13 & 1.1 & 0302 \\
NGC\,4321                    & 12 22 54.9  & $\:\,$15 49 20 &  1570 &  58 & 1100,2100 & 12 & 1.1 & 0302 \\
NGC\,4414                    & 12 26 27.1  & $\:\,$31 13 24 &   700 &  68 &  230,1250 & 17 & 1.1 & 0302 \\
NGC\,4490                    & 12 30 36.1  & $\:\,$41 38 34 &   560 &  78 &   60,1100 & 17 & 1.1 & 0302 \\
NGC\,4501                    & 12 31 59.1  & $\:\,$14 25 14 &  2278 &  55 & 1800,2800 & 10 & 1.1 & 0103 \\
NGC\,4527                    & 12 34 08.5  & $\:\,$02 39 10 &  1735 &  64 & 1240,2260 & 15 & 1.1 & 0302 \\
NGC\,4631                    & 12 42 07.9  & $\:\,$32 32 26 &   600 & 120 &  200,1100 &  7 & 4.2 & 0701 \\
NGC\,4666                    & 12 45 08.7  &--00 27 41      &  1500 &  77 & 1050,2050 & 10 & 1.1 & 0103 \\
NGC\,4736                    & 12 50 53.0  & $\:\,$41 07 13 &   300 & 105 & --150,820 &  5 & 4.2 & 0701 \\
NGC\,4826                    & 12 56 43.7  & $\:\,$21 40 52 &   400 &  76 &  --80,940 & 11 & 1.1 & 0103 \\
NGC\,5005                    & 13 10 56.3  & $\:\,$37 03 33 &   950 &  59 &  450 1470 & 13 & 1.1 & 0302 \\
NGC\,5055                    & 13 15 49.3  & $\:\,$42 01 49 &   500 & 101 &   30,1030 & 12 & 1.1 & 0302 \\
{\it{NGC\,5194}}\,$^{\rm d)}$& 13 29 52.7  & $\:\,$47 11 42 &   450 & 123 &           &    &     &      \\
NGC\,5236                    & 13 37 00.7  &--29 51 58      &   500 & 213 &   50,1050 & 34 & 1.1 & 0302 \\
Arp\,220                     & 15 34 57.1  & $\:\,$23 30 11 &  5430 & 118 & 5220,5700 & 10 & 1.1 & 0601 \\
NGC\,6000                    & 15 49 49.6  &--29 23 11      &  2200 &  59 & 1970,2400 & 25 & 1.1 & 0601 \\
NGC\,6946                    & 20 34 52.4  & $\:\,$60 09 14 &    50 & 128 & --170,300 & 11 & 1.1 & 0601 \\
NGC\,7331                    & 22 37 04.0  & $\:\,$34 24 56 &   820 &  82 &  600,1070 & 14 & 1.1 & 0601 \\
                             &             &                &       &     &  330,1070 & 12 & 1.1 & 0302 \\
\hline
\end{tabular}
\end{center}
a) Source names in italics: previously detected sources; bold-faced: newly detected sources \\
b) Flux densities are taken from the IRAS Point Source Catalog (Fullmer \& Lonsdale 1989) \\
c) Month (first two digits) and year (last two digits) \\ 
d) IC\,10: see Henkel et al. (1986); NGC\,253: Ho et al. (1987); NGC\,1068: Claussen et al. (1984); IC\,342:
   Tarchi et al. (2002a) and Sect.\,3.5; NGC\,2146: Tarchi et al. (2002b) and Sect.\,3.5; NGC\,3034 (M\,82): 
   Claussen et al. (1984); NGC\,3079: Henkel et al. (1984), Haschick \& Baan (1985); NGC\,3556 and Arp\,299: 
   this paper; NGC\,5194 (M\,51): Ho et al. (1987) \\
\end{scriptsize}
\end{table*}

\section{Observations}

The target sources of the two samples were measured in the $6_{16} - 5_{23}$ line of H$_2$O (rest frequency: 
22.23508\,GHz) with the 100-m telescope of the MPIfR at Effelsberg on various occasions between June 2001 and 
April 2004. The full width to half power beamwidth was $\sim$40\arcsec\ and the pointing accuracy was in most 
cases better than 10\arcsec\ (see also Sect.\,4.2.3). A dual channel HEMT receiver provided system 
temperatures of 130--180\,K on a main beam brightness temperature scale. The observations were carried out in 
a dual beam switching mode with a beam throw of 2\arcmin\ and a switching frequency of $\sim$1\,Hz. The 
autocorrelator backend was split into eight bands of width 40 or 80\,MHz and 512 or 256 channels each that 
could individually be shifted in frequency by up to $\pm$250\,MHz relative to the recessional velocity of the 
galaxy. This yielded channel spacings of $\sim$1 or $\sim$4\,km\,s$^{-1}$. A few spectra were also taken with 
two bands of 20\,MHz and 4096 channels each. The resulting channel spacing was then $\sim$0.07\,km\,s$^{-1}$. 
Flux calibration was obtained by measurements of W3(OH) (for the flux, see Mauersberger et al. 1988). Gain 
variations as a function of elevation were taken into account (see Eq.\,1 of Gallimore et al. 2001) and the 
1$\sigma$ flux calibration error is expected to not exceed $\pm$10\%.

\section{Results}

From the jet-maser sample, we have detected two new megamasers, Mrk~1066 and Mrk~34. The FIR-maser sample also 
yields two new detections, a megamaser (Arp~299) and a kilomaser (NGC~3556). Line profiles are shown in 
Figs.\,\ref{mrk1066a}--\ref{arp299a}. Line parameters including recessional velocity and (isotropic) H$_2$O 
luminosity are given in Table 3. Properties of the detected galaxies are discussed below.

\subsection{Mrk~1066 (UGC~2456)}

Mrk~1066 is a FIR luminous ($L_{\rm FIR}$$\sim$7$\times$10$^{10}$\,L$_{\odot}$) SB0+ galaxy, containing a 
double nucleus (e.g. Gimeno et al. 2004). It is one of the few early-type galaxies that have been detected in 
CO (see Henkel \& Wiklind 1997; note that their FIR luminosity (their Table II) is too low). Its systemic velocity 
is 3605\,km\,s$^{-1}$ (see Table 1), corresponding to a distance of $\sim$50\,Mpc. The inclination angle is 
42\degr\ (Whittle 1992). Hubble Space Telescope (HST) imaging of the nuclear region (Bower et al. 1995) shows a 
jet-like feature in a narrow-band image which includes [O{\sc iii}] and H$\beta$. The distribution is bipolar, 
oriented at $\sim$315\degr, and extending to an angular radius of $\sim$1\ffas5, with emission from the north-western 
side being dominant. In H$\alpha$ and [N{\sc ii}], the jet is equally prominent on both sides of the nucleus. 
The 3.6\,cm radio continuum emission (Nagar et al. 1999) is extended along the same axis over $\sim$2\ffas5. 

There is a strong narrow maser spike at 3636\,km\,s$^{-1}$ with a full width to half maximum linewidth of less 
than 2\,km\,s$^{-1}$ (Figs.\,\ref{mrk1066a} and \ref{mrk1066b}) and a peak flux density of 80\,mJy. The spike 
becomes narrower between March 7 and 10, 2002, appears to be unresolved in frequency in May and reaches only 
$\sim$50\,mJy in September. It seems that a gradual decrease of the linewidth is finally accompanied by a decrease 
in peak flux density. At a distance of $\sim$50\,Mpc, isotropic luminosities reach $\sim$10\,L$_{\odot}$. A second 
much wider component, at $\sim$3550\,km\,s$^{-1}$, has a peak flux density of 10--20\,mJy and an isotropic luminosity 
of $\sim$40\,L$_{\odot}$.

\subsection{Mrk\,34}

Mrk~34 (IRAS~10309+6017), another luminous ($\sim$10$^{11}$\,L$_{\odot}$) FIR source, is a distant Seyfert 2 
galaxy ($z$=0.0505, $D$$\sim$200\,Mpc; Falcke et al. 1998). The optical galaxy is characterized as having an 
inclination angle of 57\degr\ in Whittle (1992), though a second generation Digital Sky Survey (DSS) image shows 
the galaxy to be compact, with poorly defined outer isophotes (Nagar \& Wilson 1999). The radio emission has an 
extended structure ($\sim$2\ffas5; Ulvestad \& Wilson 1984), and strong evidence for an interaction between 
the radio jet and NLR clouds has been found by Falcke et al. (1998).

Mrk~34 is one of the most distant and most luminous H$_2$O megamasers ever detected. The maser shows two or 
three distinct spectral features (Figs.\,\ref{mrk34a} and \ref{mrk34b}). One is centered at a velocity of 
$\sim$14840\,km\,s$^{-1}$, another at $\sim$15770\,km\,s$^{-1}$, and a third one is tentatively seen at 
$\sim$14665\,km\,s$^{-1}$. Peak flux densities are up to 10\,mJy and total isotropic luminosities are 
$\sim$1000\,L$_{\odot}$.

\subsection{NGC~3556 (M~108)}

NGC~3556 is {\bf an} edge-on spiral galaxy located at a distance of $\sim$12\,Mpc. Its FIR luminosity, 
$L_{\rm FIR}$ $\sim$ 10$^{10}$\,L$_{\odot}$, is similar to that of the Milky Way. Radio continuum, \HI\ and 
X-ray data indicate a violent disk halo interaction, including a prominent radio halo (e.g. de Bruyn \& Hummel 
1979), large \HI\ extensions possibly delineating expanding supershells (King \& Irwin 1997), compact radio 
continuum sources, likely representing supernova remnants (Irwin et al. 2000), and extraplanar diffuse X-ray 
emission (Wang et al. 2003). $^{12}$CO and HCN observations (Gao \& Solomon 2004) indicate a substantial
molecular gas content. No OH maser was detected (Unger et al. 1986).

\begin{figure}[ht]
\vspace{-1.0cm}
\hspace{0.9cm}
\includegraphics[bb=58 39 548 571, angle=-90, width=15cm]{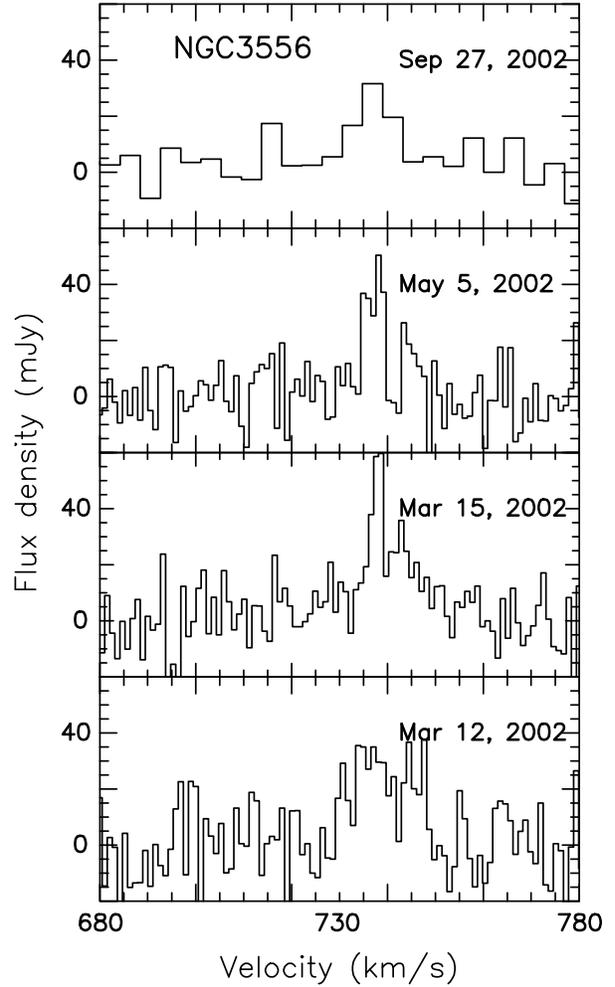}
\vspace{0.5cm}
\caption[fig5]{\footnotesize{H$_2$O kilomaser spectra toward NCG~3556 ($\alpha_{2000}$ = 11$^{\rm h}$ 11$^{\rm m}$ 31\ffs2, 
$\delta_{2000}$ = 55$^{\circ}$ 40\arcmin\ 25\arcsec). Channel spacings are 4.2\,km\,s$^{-1}$ (upper panel) and 1.06\,km\,s$^{-1}$. 
c$z$ = 700\,km\,s$^{-1}$ (NED). $V_{\rm LSR}$--$V_{\rm HEL}$ = +5.90\,km\,s$^{-1}$.}
\label{ngc3556}}
\end{figure}

With a peak flux density of 20--40\,mJy, the H$_2$O maser has an isotropic luminosity of $\sim$1\,L$_{\odot}$. 
Only one velocity component is seen. The profiles are shown in Fig.\,\ref{ngc3556}.

\begin{table*}
\label{fluxes}
\begin{scriptsize}
\begin{center}
\caption{Line parameters of newly detected masers}
\begin{tabular}{lcccrrcccl}
\hline
Source & Adopted          & Epoch & $\int{S{\rm d}V}^{\rm a)}$    & $V_{\rm LSR,opt}^{\rm a,b)}$ & $\Delta V_{1/2}^{\rm a)}$ &
         Channel          & rms   & log\,$L_{\rm H_2O}^{\rm d)}$  & Comments \\
       & Distance         &       &                               &                              &                           &
         Width$^{\rm c)}$ &       &                               &          \\  
       & (Mpc)            & (2002)& (mJy km\,s$^{-1}$)            & (km\,s$^{-1}$)               & (km\,s$^{-1}$)            &
         (km\,s$^{-1}$)   & (mJy) & (L$_{\odot}$)                 &          \\  
\\
\hline 
Mrk\,1066          & 48&Mar 07&  696$\pm$61  & 3547.2$\pm$1.4 &  38.7$\pm$04.6 & 1.08 & 5.3&    1.6     & see Fig.\,1 \\
                   &   &      &  169$\pm$12  & 3635.9$\pm$0.1 &   2.1$\pm$00.2 & 1.08 & 5.3&    1.0     & see Fig.\,1 \\
                   &   &      &  171$\pm$17  & 3635.9$\pm$0.1 &   1.9$\pm$00.2 & 0.07 &29.7&    1.0     & see Fig.\,2 \\
                   &   &Mar 10&  822$\pm$59  & 3547.8$\pm$1.3 &  38.3$\pm$03.4 & 1.08 & 5.6&    1.6     & see Fig.\,1 \\
                   &   &      &  193$\pm$13  & 3636.2$\pm$0.1 &   2.1$\pm$00.2 & 1.08 & 5.6&    1.0     & see Fig.\,1 \\
                   &   &      &  157$\pm$16  & 3636.1$\pm$0.1 &   1.2$\pm$00.2 & 0.07 &34.9&    0.9     & see Fig.\,2 \\
                   &   &May 03&  566$\pm$16  & 3552.8$\pm$7.7 &  46.2$\pm$16.9 & 1.08 &15.0&    1.5     & see Fig.\,1 \\
                   &   &      &  166$\pm$33  & 3635.9$\pm$0.2 &   1.7$\pm$00.4 & 1.08 &15.0&    0.9     & see Fig.\,1 \\
                   &   &Sep 27&  143$\pm$31  & 3635.2$\pm$0.2 &   2.0$\pm$00.5 & 1.08 &13.7&    0.9     & see Fig.\,1 \\
                   &   &      &              &                &                &           &            &             \\
Mrk\,34$^{\rm e)}$ &205&Mar 08&   46$\pm$14  &14661.2$\pm$0.4 &   2.1$\pm$00.8 & 1.16 & 5.8&    1.6     & see Fig.\,3 \\
                   &   &      &  354$\pm$52  &14837.6$\pm$2.3 &  31.9$\pm$05.3 & 1.16 & 5.8&    2.5     & see Fig.\,3 \\
                   &   &      &  417$\pm$78  &15771.7$\pm$8.7 &  88.2$\pm$16.2 & 9.29 & 1.9&    2.6     & see Fig.\,4 \\
                   &   &Mar 17&  335$\pm$57  &14840.5$\pm$2.9 &  36.5$\pm$07.8 & 1.16 & 5.5&    2.5     & see Fig.\,3 \\
                   &   &May 07&   77$\pm$24  &14674.9$\pm$0.8 &   5.0$\pm$01.9 & 1.16 & 6.6&    1.9     & see Fig.\,3 \\
                   &   &      &  323$\pm$62  &14838.7$\pm$2.9 &  31.2$\pm$07.6 & 1.16 & 6.6&    2.5     & see Fig.\,3 \\
                   &   &      &  995$\pm$122 &15754.9$\pm$8.0 & 130.4$\pm$18.4 & 9.30 & 2.3&    3.0     & see Fig.\,4 \\
                   &   &May 09&   67$\pm$25  &14658.2$\pm$0.8 &   4.7$\pm$02.2 & 4.67 & 3.4&    1.8     & see Fig.\,3 \\
                   &   &      &  559$\pm$81  &14837.4$\pm$3.1 &  43.9$\pm$08.2 & 4.67 & 3.4&    2.7     & see Fig.\,3 \\
                   &   &      &  607$\pm$92  &15787.2$\pm$9.6 & 115.5$\pm$19.9 &18.67 & 1.4&    2.8     & see Fig.\,4 \\
                   &   &      &  117$\pm$32  &15738.2$\pm$1.1 &   6.9$\pm$03.0 & 4.67 & 3.2&    2.1     & see Fig.\,4 \\
                   &   &      &  394$\pm$87  &15803.1$\pm$4.2 &  47.8$\pm$15.0 & 4.67 & 3.2&    2.6     & see Fig.\,4 \\
                   &   &      &              &                &                &           &            &             \\
NGC\,3556          & 12&Mar 12&  505$\pm$72  &  738.5$\pm$1.1 &  15.1$\pm$2.2  & 1.06 &11.6&    0.2     & see Fig.\,5 \\
                   &   &Mar 15&   99$\pm$29  &  738.0$\pm$0.3 &   2.1$\pm$0.5  & 1.06 &10.1&  --0.5     & see Fig.\,5 \\
                   &   &      &  421$\pm$78  &  740.3$\pm$1.6 &  17.9$\pm$4.8  & 1.06 &10.1&    0.1     & see Fig.\,5 \\
                   &   &May 05&  229$\pm$35  &  737.2$\pm$0.4 &   4.8$\pm$0.8  & 1.06 &10.9&  --0.1     & see Fig.\,5 \\
                   &   &Sep 27&  329$\pm$75  &  737.1$\pm$1.1 &  10.0$\pm$2.9  & 4.21 & 6.5&    0.0     & see Fig.\,5 \\
                   &   &      &              &                &                &           &            &             \\
Arp\,299           & 42&Mar 16& 6210$\pm$242 & 3101.8$\pm$05.1& 260.2$\pm$11.5 & 4.30 & 4.8&    2.4     & see Fig.\,6 \\
\\
\hline
\end{tabular}
\end{center}
a) Obtained from Gaussian fits \\
b) See the caption to Fig.\,\ref{mrk1066a} \\
c) Channel width of the spectra used for the Gaussian fits. Some of the 
   spectra shown in Figs.\,\ref{mrk1066a}--\ref{arp299b} are smoothed \\
d) $L_{\rm H_2O}$/[L$_{\odot}$] = 0.023 $\times$ $\int{S {\rm d}V}$/[Jy\,km\,s$^{-1}$] $\times$ $D^2$/[Mpc$^2$] \\
e) To the observations of the high velocity feature (15787\,km\,s$^{-1}$) from May 9: The first Gaussian 
   fit refers to the smoothed profile shown in Fig.\,\ref{mrk34b}; the latter two component fit refers to the 
   unsmoothed spectrum with a channel spacing of 4.7\,km\,s$^{-1}$ \\
\end{scriptsize}
\end{table*}

\subsection{Arp~299 (Mrk~171)}

Arp~299 is a merging system at $D$$\sim$40\,Mpc, composed of two main sources, IC~694 and NGC~3690 (for an 
alternative nomenclature, see Sect.\,4.2.3), that are separated by 22\arcsec\ in east-west direction (e.g. Sargent 
\& Scoville 1991). A FIR luminosity of several 10$^{11}$\,L$_{\odot}$ (Casoli et al. 1992) places Arp~299 near the 
boundary between luminous (LIRGs) and ultraluminous (ULIRGs) infrared galaxies. Supporting the merging scenario, 
two highly extended \HI\ tails have been identified by Hibbard \& Yun (1999). Radio and infrared observations 
reveal three main regions of activity (e.g. Gehrz et al. 1983; Aalto et al. 1997; Casoli et al. 1999), the nuclear 
regions of IC~694 and NGC~3690 and an interface where IC~694 and NGC~3690 overlap. NGC~3690 contains a deeply 
enshrouded active galactic nucleus (AGN), while the situation with respect to the similarly obscured nuclear 
region of IC~694 is less clear (e.g. Della Ceca et al. 2002; Ballo et al. 2004; Gallais et al. 2004). $^{12}$CO 
and HCN $J$=1--0 emission line peaks are strongest toward these most active regions, indicating the presence of 
large amounts of molecular gas. The positions of strongest $^{13}$CO $J$=1$-$0 line emission are, however, 
displaced from these hotspots (Aalto et al. 1997).

In Arp~299, water maser profiles are extremely broad ($\sim$200\,km\,s$^{-1}$), with peak flux densities of 
30\,mJy (Fig.\,\ref{arp299a}). Adopting a distance of 42\,Mpc, the total isotropic luminosity is 
$\sim$250\,L$_{\odot}$, placing the object among the more luminous H$_2$O megamaser sources. The maser line 
is centered at a velocity of 3100\,km\,s$^{-1}$, i.e. close to the systemic velocity of the entire complex of 
sources constituting Arp~299.

\subsection{IC~342 and NGC~2146}

We also observed the previously detected H$_2$O maser sources IC~342 and NGC~2146. IC~342 was not detected in 
June 2001 and March 2002, indicating that the flaring component observed at $V_{\rm LSR}$ $\sim$ 16\,km\,s$^{-1}$ 
(Tarchi et al. 2002a) has been quiescent since June 2001. Spectra from NGC~2146, obtained in March 2002, show no 
significant variations with respect to profiles observed two years earlier (see Tarchi et al. 2002b).

\section{Discussion}

As indicated in Sect.\,1, the surveys presented here have been targeted to detect two classes of 
extragalactic water masers, `jet-masers' and `FIR-masers'.

\subsection{The jet-maser sample}

Jet-masers provide insight into the interaction of nuclear jets with dense warm molecular gas in the central 
parsecs of galaxies. All jet-masers known to date arise from the innermost regions of active galaxies and 
yield important information about the evolution of jets and their hotspots. If continuum emission from the core 
of the radio source is responsible for variations in maser intensity, monitoring of continuum and line emission 
can provide estimates, through reverberation mapping, of the speed of the material in the jet, particularly in 
sources where the jet appears to lie close to the plane of the sky. If, on the other hand, the continuum flare 
is caused by the brightening of the hotspot or working surface in the jet as it impacts a denser molecular cloud, 
then the onset of the continuum and maser flares should be nearly simultaneous (Peck et al. 2003).

\begin{figure}[ht]
\vspace{-1.0cm}
\hspace{0.9cm}
\includegraphics[bb=58 39 548 571, angle=-90, width=15cm]{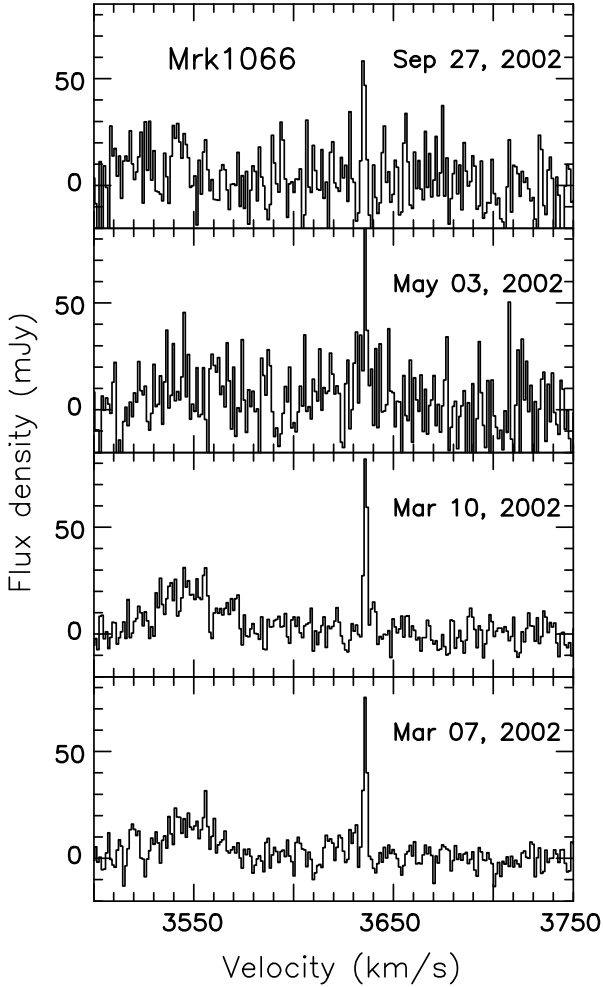}
\vspace{0.4cm}
\caption[fig1]{\footnotesize{22\,GHz H$_2$O megamaser profiles toward Mrk\,1066 with a channel spacing of 1.08\,km\,s$^{-1}$. 
$\alpha_{2000}$ = 02$^{\rm h}$ 59$^{\rm m}$ 58\ffs6, $\delta_{2000}$ = 36$^{\circ}$ 49\arcmin\ 14\arcsec. Velocity scales are 
with respect to the Local Standard of Rest (LSR) and use the optical convention that is equivalent to c$z$. $V_{\rm sys}$ = 
c$z_{\rm sys}$ = 3605\,km\,s$^{-1}$ (NASA/IPAC Extragalactic Database (NED)). $V_{\rm LSR}$--$V_{\rm HEL}$ = --3.55\,km\,s$^{-1}$.}
\label{mrk1066a}}
\end{figure}

\begin{figure}[ht]
\vspace{-6.0cm}
\hspace{0.9cm}
\includegraphics[bb=58 39 548 571, angle=-90, width=15cm]{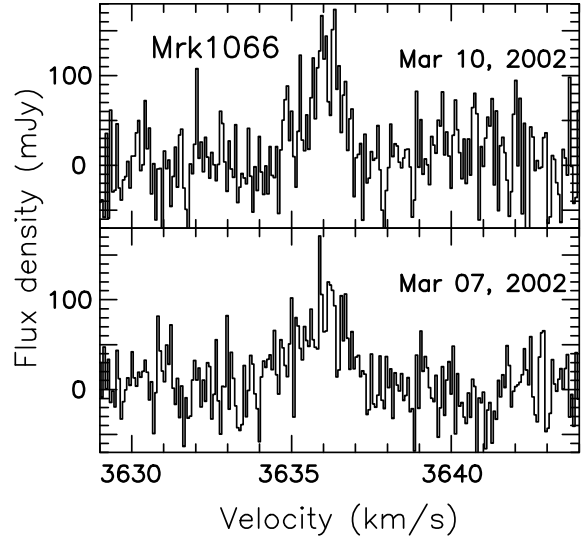}
\vspace{0.5cm}
\caption[fig2]{\footnotesize{High spectral resolution profiles of the narrow maser spike in Mrk\,1066 (see Fig.\,\ref{mrk1066a}). 
The channel spacing is 0.067\,km\,s$^{-1}$.}
\label{mrk1066b}}
\end{figure}

In view of these implications we need to investigate the true nature of the megamasers detected in Mrk~1066 and Mrk~34. 
Are these really jet-masers as suggested by the selection criteria (Sect.\,1)? While only VLBI observations can provide 
a definite answer, a detailed look at well studied jet-masers and the maser sources recently discovered by Braatz et al. 
(2004) can provide relevant information. 

The four jet-maser sources known prior to this survey are NGC~1068 (Gallimore et al. 1996), the Circinus galaxy 
(Greenhill et al. 2001, 2003a), NGC~1052 (Claussen et al. 1998) and Mrk~348 (Peck et al. 2003). The first two sources 
show both maser emission from a circumnuclear disk {\it and} emission arising along the edges of an ionization cone 
or outflow in the jet. In NGC\,1068, the jet-maser velocities are blue-shifted by $\sim$160\,km\,s$^{-1}$ from 
systemic and the feature is broad (FWHP (full width to half power) linewidth $\sim$ 60\,km\,s$^{-1}$; e.g. Gallimore 
et al. 2001). In Circinus, both red- and blue-shifted features are seen at velocities up to 160\,km\,s$^{-1}$ from 
systemic (Greenhill et al. 2003a). In NGC~1052 and Mrk~348, all the maser emission may arise along the jet (Claussen 
et al. 1998; Peck et al. 2003). As in NGC~1068, this is accompanied by relatively large linewidths ($\sim$90 and 
$\sim$130\,km\,s$^{-1}$) and significant shifts relative to the systemic velocity ($\sim$ +150--200 and 
+130\,km\,s$^{-1}$), respectively. 

To summarize, jet-maser features tend to be broader (a few 10\,km\,s$^{-1}$) than those typically seen in disk-maser 
sources like NGC\,4258 (a few km\,s$^{-1}$) and are usually displaced from the systemic velocity. In Mrk~1066, it is the 
component at c$z$$\sim$3550\,km\,s$^{-1}$ that shows the properties expected in the case of a nuclear jet-type H$_2$O maser 
(see Fig.\,\ref{mrk1066a}). The intense narrow spike near the systemic velocity would have a different origin. We also note, 
however, that the broad blue- and the narrow red-shifted features bracket the systemic velocity (3605\,km\,s$^{-1}$). Thus 
a masing disk like in NGC~4258 cannot be excluded.

\begin{figure}[ht]
\hspace{0.9cm}
\includegraphics[bb=58 39 548 571, angle=-90, width=15cm]{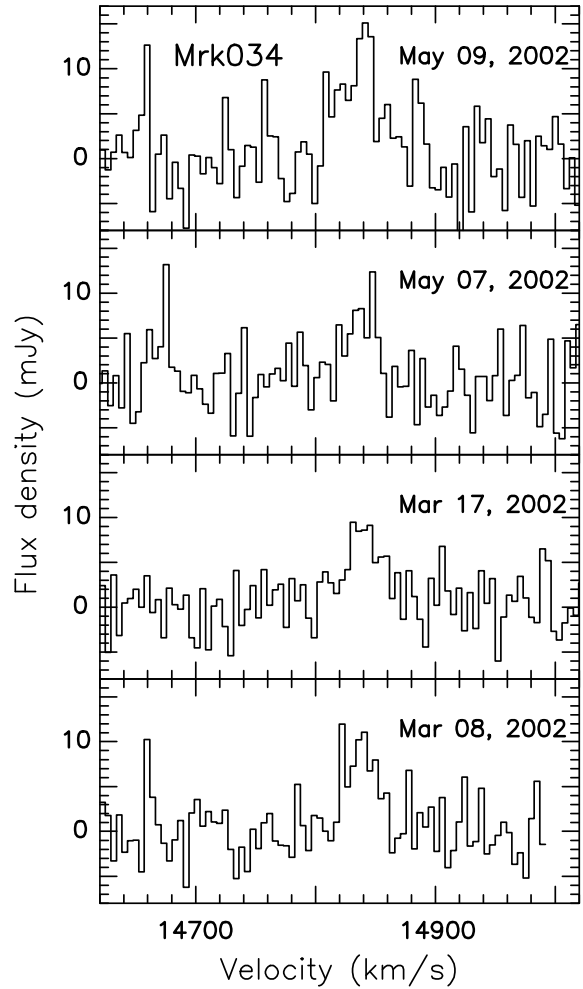}
\vspace{0.5cm}
\caption[fig3]{\footnotesize{Low velocity H$_2$O megamaser profiles toward Mrk\,34 with a channel spacing of 4.65\,km\,s$^{-1}$. 
$\alpha_{2000}$ = 10$^{\rm h}$ 34$^{\rm m}$ 08\ffs6, $\delta_{2000}$ = 60$^{\circ}$ 01\arcmin\ 52\arcsec. c$z_{\rm sys}$ = 
15145\,km\,s$^{-1}$ (de Grijp et al. 1992). $V_{\rm LSR}$--$V_{\rm HEL}$ = +5.23\,km\,s$^{-1}$.} 
\label{mrk34a}}
\end{figure}

\begin{figure}[ht]
\vspace{-3.0cm}
\hspace{0.9cm}
\includegraphics[bb=58 39 548 571, angle=-90, width=15cm]{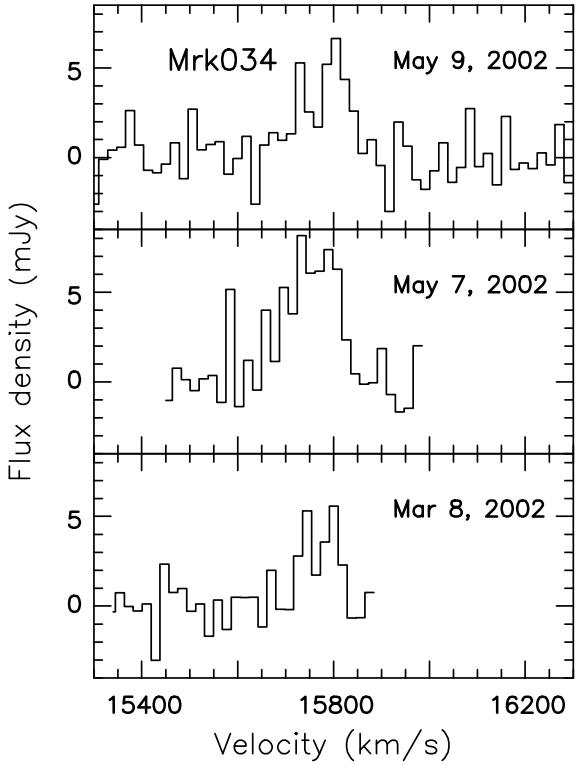}
\vspace{0.5cm}
\caption[fig3]{\footnotesize{High velocity megamaser feature toward Mrk\,34 with a channel spacing of 18.65\,km\,s$^{-1}$.}
\label{mrk34b}}
\end{figure}

Toward Mrk~34, the main components at c$z$$\sim$14840 and 15770\,km\,s$^{-1}$ are wide enough for characteristic 
jet-maser emission. However, the intrinsic weakness of the features requires smoothing which could hide individual 
narrow components that might represent a significant fraction of the maser emission. Furthermore, Figs.\,\ref{mrk34a} 
and \ref{mrk34b} show two or three velocity components that bracket the systemic velocity ($V_{\rm sys}$ = 
15145$\pm$90\,km\,s$^{-1}$; de Grijp et al. 1992). The velocity displacements may not be symmetric; the red-shifted 
emission (Fig.\,\ref{mrk34b}) appears to show a larger displacement than the blue-shifted emission (Fig.\,\ref{mrk34a})
from systemic, which would argue against the possibility of a circumnuclear disk. However, the uncertainty in 
c$z_{\rm sys}$ is large so that an accretion `disk-maser' scenario is also possible. Among the three `jet-maser'
sources detected by Braatz et al. (2004; see also Table 1), Mrk~1157 (NGC~591), Mrk~3 and NGC~4151, the first one also 
shows a profile reminiscent of a disk-maser source. We thus conclude that our jet-maser sample does not provide 
exclusively jet-maser sources. Having selected sources with jets that appear to be oriented close to the plane of
the sky (Sect.\,1), this is apparently also an excellent selection criterion to find disk-masers that are characterized
by nuclear disks viewed edge-on. Disk-masers may constitute a significant fraction of the newly discovered `jet-maser' 
sources and some of these may even show both signatures (like NGC~1068) of nuclear activity.

Including all sources that fulfill the selection criteria of our jet-maser sample (Table 1), the detection rate 
becomes an (almost incredible) 7/14 or 50\%. This is the first survey undertaken to look specifically for jet-masers. 
The number of sources and detections is still too small for a detailed statistical analysis. While it remains to be
seen whether the masers are jet- or disk-masers, the unprecedented success rate suggests that both types of masers 
have been found and that a tilt of $>$55\degr\ between nuclear and large scale disk is a highly favorable configuration 
for the occurrence of H$_2$O masers in Seyfert galaxies.

\subsection{The FIR-maser sample}

\subsubsection{Previously detected masers}

FIR emission commonly arises from dust grains heated by newly formed stars. In the Milky Way, 22\,GHz H$_2$O masers are 
associated with sites of (mostly massive) star formation. Therefore our sample of FIR bright galaxies (Table 2) is a 
suitable tool to detect extragalactic H$_2$O masers associated with young massive stars. Such masers have the potential 
to pinpoint the location of prominent star forming regions and to estimate their distance through complementary measurements 
of proper motion and radial velocity (e.g. Greenhill et al. 1993). Monitoring such masers and determining their three 
dimensional velocity vectors allows us to derive the gravitational potential of galaxies or groups of galaxies and to 
improve our understanding of the evolution of such groups with time (for the Local Group, see Brunthaler et al. 2002). 

Including the early part of our survey (Tarchi et al. 2002a,b), we detected with IC~342, NGC~2146, NGC~3556 and Arp~299 
four new H$_2$O masers in a total of 45 sources (see Table 2). The new detections are a consequence of higher sensitivity 
(1$\sigma$ noise levels of $\sim$ 10\,mJy for a 1\,km\,s$^{-1}$ channel), highly improved baselines and luck (in the case 
of the short-lived flare observed toward IC~342). Including all previously detected sources in the complete sample shown 
in Table 2, we find a detection rate of 10/45 or 22$\pm$7\%. For sources with 100$\mu$m fluxes in excess of 100\,Jy, the detection 
rate becomes even higher: 7/19 or 37$\pm$14\%. Detection rates for the jet-maser and the FIR-maser samples lie far above the 
corresponding rates deduced from other carefully selected samples (see e.g. Henkel et al. 1984, 1986, 1998; Haschick \& Baan 
1985; Braatz et al. 1996; Greenhill et al. 2002). 

To find out why the FIR-maser sample contains numerous 22\,GHz H$_2$O maser sources and to elucidate the nature of the 
sources in NGC~3556 and Arp~299, we have to classify the properties of the previously studied masers of this sample. Two 
of the sources, those in NGC~1068 and NGC~3079, are luminous megamasers (e.g. Gallimore et al. 2001; Trotter et al. 1998). 
The weaker `kilomasers' in IC~10, IC~342, NGC~2146 and NGC~3034 (M~82) are associated with sites of massive star formation 
(Argon et al. 1994; Baudry \& Brouillet 1996; Tarchi et al. 2002a,b). There are also two known weak nuclear kilomasers, 
in NGC~5194 (M~51) (Hagiwara et al. 2001b) and in NGC~253 (Henkel et al. 2004). Whether they are related to the nearby AGN or 
to star formation remains, however, unclear.

\subsubsection{NGC~3556}

The position of the H$_2$O kilomaser in NGC~3556 is not yet accurately known. From the relative number of nuclear versus 
non-nuclear masers of similar luminosity, the most likely interpretation is an association with a site of massive star 
formation. In view of NGC~253 and NGC~5194, however, there is a small chance for a nuclear maser in NGC~3556.

\begin{figure}[ht]
\vspace{-9.1cm}
\hspace{0.9cm}
\includegraphics[bb=58 39 548 571, angle=-90, width=15cm]{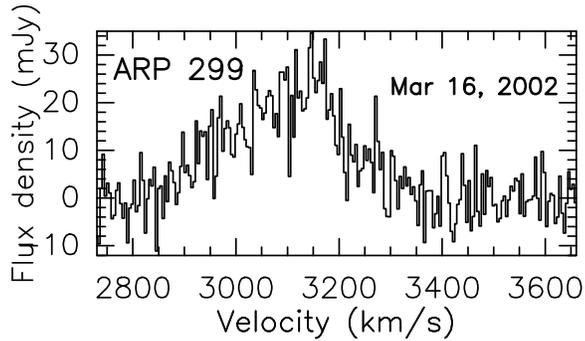}
\vspace{0.5cm}
\caption[fig10]{\footnotesize{H$_2$O megamaser profile toward Arp\,299 ($\alpha_{2000}$ = 11$^{\rm h}$ 28$^{\rm m}$ 31\ffs9, 
$\delta_{2000}$ = 58$^{\circ}$ 33\arcmin\ 45\arcsec). The channel spacing is 4.3\,km\,s$^{-1}$. $V_{\rm LSR}$--$V_{\rm HEL}$ 
= +6.97\,km\,s$^{-1}$. For radial velocities, see Sect.\,4.2.3.}
\label{arp299a}}
\end{figure}

\subsubsection{Arp~299}

\begin{figure*}[ht]
\hspace{0.9cm}
\includegraphics[bb=58 39 548 571, angle=-90, width=12cm]{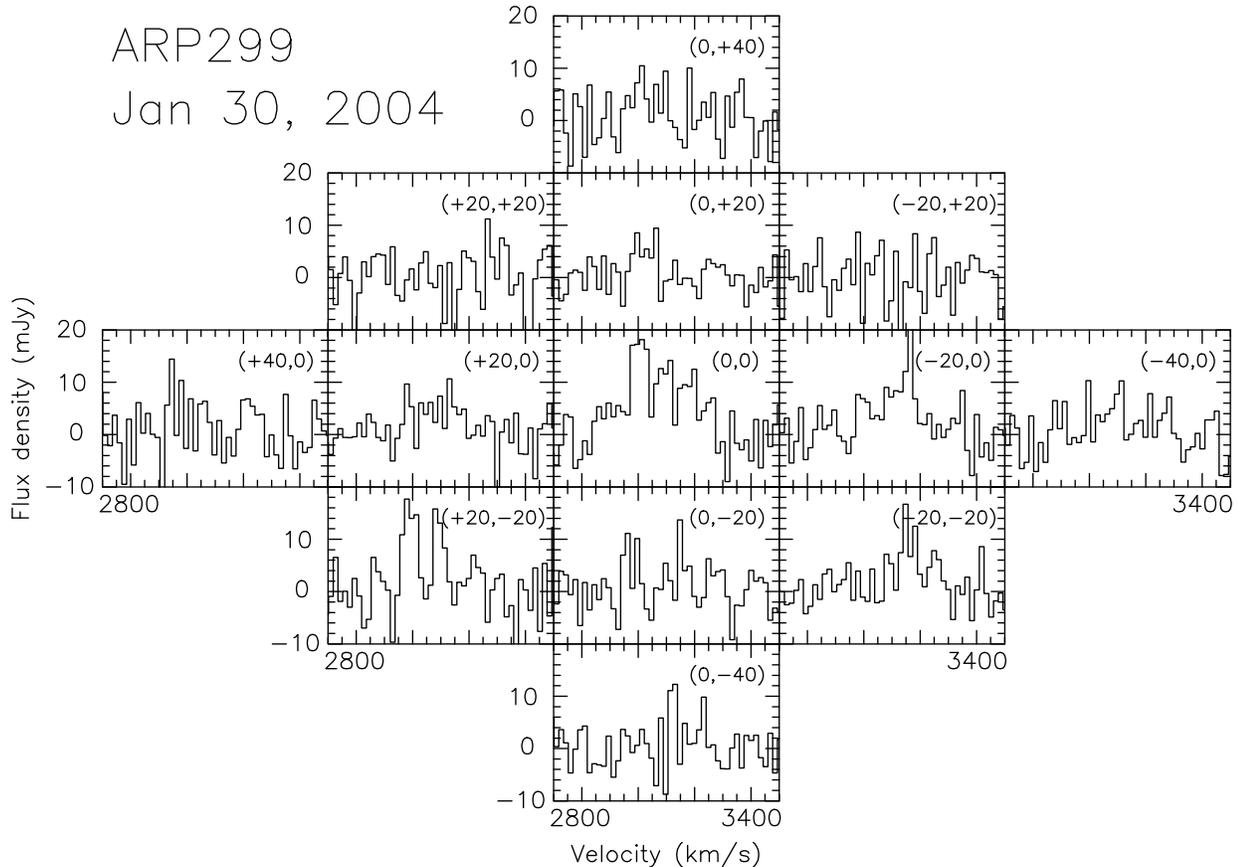}
\caption[fig4]{\footnotesize{H$_2$O megamaser profiles toward Arp\,299, taken on Jan 30, 2004, during a night with excellent 
pointing conditions. The reference position is $\alpha_{2000}$ = 11$^{\rm h}$ 28$^{\rm m}$ 31\ffs9, $\delta_{2000}$ = 
58$^{\circ}$ 33\arcmin\ 45\arcsec. Offsets in arcsec ($\Delta \alpha$, $\Delta \delta$) are given in the upper right corner 
of each box. Spectra with offsets $\Delta \alpha$  = --20\arcsec\ trace emission from the `overlap' region (north) and 
NGC\,3690 (south). Spectra with $\Delta \alpha$ = +20\arcsec\ are sensitive to emission from IC\,694 (see Sect.\,4.2.3). 
The channel spacing is 16.8\,km\,s$^{-1}$.}
\label{arp299b}}
\end{figure*}

As indicated in Sect.\,3.4, the merging system Arp~299 is composed of two galaxies, NGC~3690 in the west, IC~694 in the 
east, and a star-bursting interface or overlap region 10\arcsec\ north of NGC~3690\footnote{According to NED, the entire 
merger forms NGC~3690, while IC~694 is a less prominent galaxy $\sim$1$'$ to the northwest. The latter target was observed 
by Braatz et al. (2004) in the H$_2$O line. A 1$\sigma$ noise limit of 2.4\,mJy per 0.3\,km\,s$^{-1}$ channel was obtained. 
Here we follow the more traditional nomenclature that is commonly used in the literature.}. Since NGC~3690 and IC~694 are 
only half a beam size (20\arcsec) apart in our observations and since the separation between NGC~3690 and the overlap region 
is even smaller, extremely good pointing conditions were needed to map the region. Three maps were made. Fig.\,\ref{arp299b} 
shows the most accurate (pointing accuracy $\pm$4\arcsec) and extended, albeit also the most noisy one. In spite of the rather 
low signal-to-noise ratios we note that (1) H$_2$O emission may originate from more than one hotspot; (2) one of the potential 
sources, the one in the east, is close to IC~694, where an OH megamaser was already reported (Baan \& Haschick 1990); (3) there 
appears to be a western peak of emission that is located near the center of the second dominant galaxy of the system, NGC~3690; 
(4) a broader feature near the center is likely caused by blending of the two main hotspots associated with IC~694 and NGC~3690; 
(5) the vigorously star forming overlap region appears to be devoid of H$_2$O megamaser emission.  

When we compare these results with the CO velocity field observed by Casoli et al. (1999), we find that the H$_2$O velocity 
of the eastern peak, 2980\,km\,s$^{-1}$, is consistent with the CO velocity of the south-eastern part of IC~694 (i.e. offset 
w.r.t. the nucleus of IC~694 that has a velocity of c$z$ = 3110\,km\,s$^{-1}$). Although the location of the western hotspot 
is close to the core of NGC~3690, the velocities of the maser emission, the CO lines and the systemic velocity of NGC~3690 
do not match perfectly ($\sim$3100\,km\,s$^{-1}$ from CO, c$z_{\rm sys}$ = 3121\,km\,s$^{-1}$ (NASA/IPAC Extragalactic
Database (NED)), versus $\sim$3150\,km\,s$^{-1}$ from H$_2$O). Interestingly, velocities near 3150\,km\,s$^{-1}$ as 
seen in H$_2$O are consistent with those of the overlap region. 

A comparison of the profile shown in Fig.\,\ref{arp299a} with the central one in Fig.\,\ref{arp299b} suggests a slight offset 
in position (a few arcsec in east-west direction) and weaker peak emission in the latter case. This is within the uncertainties
of pointing and calibration, but maser variability can also explain the differences. 

With the H$_2$O emission likely originating from IC~694 and NGC~3690, Arp~299 is the fourth extragalactic system beyond the 
Magellanic Clouds that is known to exhibit $\lambda$ = 18\,cm OH and $\lambda$ = 1.3\,cm H$_2$O maser emission (in NGC~253, 
NGC~1068 and M~82, such masers are also observed; see Weliachew et al. 1984; Turner 1985; Baudry \& Brouillet 1996; Gallimore 
et al. 1996; Henkel et al. 2004). In these other galaxies, however, either H$_2$O or OH or both lines only reach kilomaser 
luminosities. IC~694 may thus be the first known galaxy with both an OH and an H$_2$O megamaser (for OH, see Baan \& Haschick 
1990). The global OH and H$_2$O line profiles appear to be similar, except for a weak OH feature at $\sim$3500\,km\,s$^{-1}$
that is not seen in H$_2$O.

Arp~299 is the second most luminous FIR source with a known H$_2$O megamaser. While OH megamasers are closely associated 
with ultraluminous infrared galaxies (ULIRGs; see e.g. Darling \& Giovanelli 2002a), the only ULIRG with a luminous H$_2$O 
maser was so far NGC~6240 (Hagiwara et al. 2002, 2003a; Nakai et al. 2002; Braatz et al. 2003). In accordance with the 
observed anticorrelation between the occurence of OH and H$_2$O megamasers (OH megamasers may arise from low density
molecular gas, while H$_2$O megamasers originate from gas of much higher density; e.g. Kylafis et al. 1991; Randell et al. 
1995), this is one of those ULIRGs in which no OH emission is seen. The second detection of an H$_2$O megamaser in a 
luminous FIR galaxy (although Arp~299 is not quite as luminous as NGC~6240) makes a dedicated survey of such luminous 
FIR galaxies worthwhile.

\subsubsection{Detection probabilities}

The high rate of maser detections in our sample of FIR luminous galaxies (Table 2) strongly suggests that a relationship 
between FIR flux density and maser phenomena exists, consistent with the assessment of HWB. The detection rate for masers
in accretion disks is dictated by tight geometric constraints. The cumulative maser output of a star forming region
may not be so narrowly confined. 

\begin{figure}[ht]
\vspace{-3.0cm}
\hspace{0.5cm}
\includegraphics[bb=58 39 548 571, angle=-90, width=9.2cm]{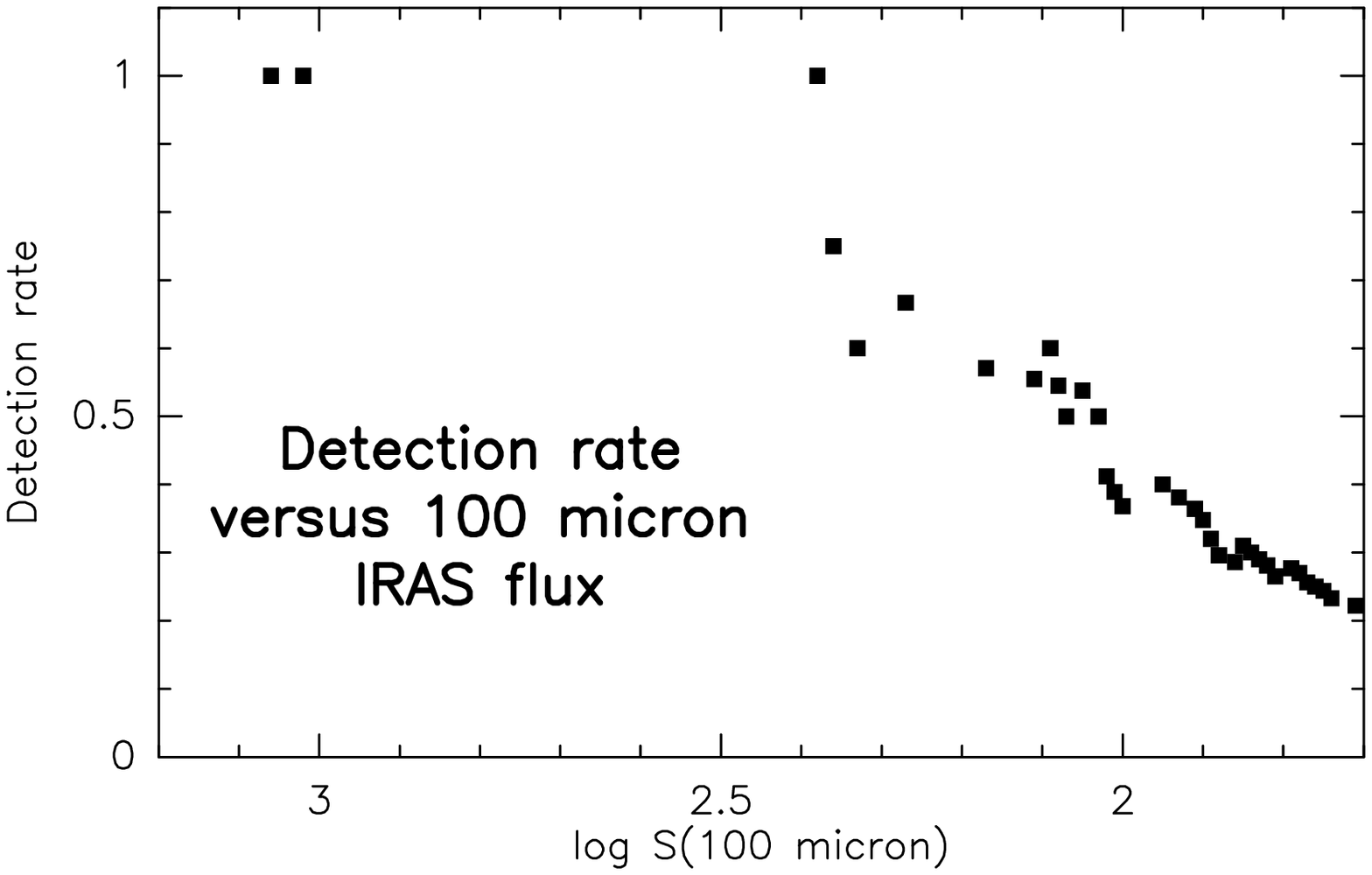}
\vspace{0.5cm}
\caption[fig8]{\footnotesize{Detection rate of the H$_2$O FIR-maser sample
(see Table 2 for the targets and Sect.\,1 for selection criteria) including
all galaxies above a given IRAS Point Source Catalog $S_{\rm 100\mu m}$ flux 
density.} 
\label{rate}}
\end{figure}

Fig.\,\ref{rate} shows the cumulative detection rate above a given 100$\mu$m IRAS Point Source Catalog flux for the parent 
galaxy. The detection rate strongly declines with decreasing FIR flux. For fluxes $\sim$1000\,Jy, 100--300\,Jy, and 50--100\,Jy, 
we find detection rates of 2/2 or 100\%, 5/17 or 29\% and 3/26 or 12\% (unlike in Fig.\,\ref{rate}, these are not
cumulative detection rates but detection rates related to their specific FIR flux density interval). Maffei~2 and NGC~5236 
(M~83) show no detectable maser emission near their nuclei but have $S_{\rm 100\mu m}$$>$200\,Jy. In view of the statistical 
properties of the sample, frequent monitoring of these sources would likely reveal H$_2$O maser emission, possibly a short-lived 
flare like that seen in IC~342 (Tarchi et al. 2002a). 

Fig.\,\ref{rate} shows a detection probability of $\sim$50\% for sources with $S_{\rm 100\mu m}$$\ga$120\,Jy. If the two 
brightest FIR sources, NGC~253 and M~82 (NGC~3034), were at $D$$\sim$10\,Mpc (i.e. three times their estimated distance), 
this would imply $S_{\rm 100\mu m}$$\sim$100\,Jy and H$_2$O peak fluxes of $\sim$5\,mJy (broad emission feature) and 
$\sim$10\,mJy (narrow emission feature), respectively (for the line profiles, see Ho et al. 1987; Baudry et al. 1994; 
Henkel et al. 2004). Thus the two sources would be just below the detection limit, consistent with the detection probability 
at the corresponding 100$\mu$m flux. 

\begin{figure}[ht]
\vspace{-0.0cm}
\hspace{0.5cm}
\includegraphics[bb=58 39 548 571, angle=-90, width=9.2cm]{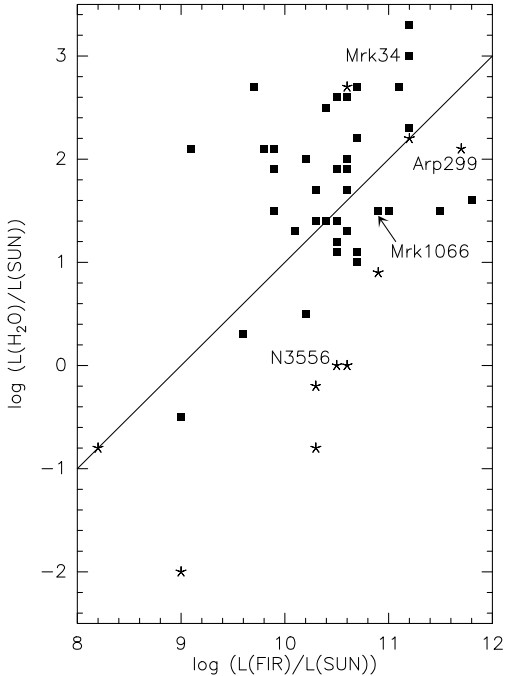}
\vspace{0.5cm}
\caption[fig9]{\footnotesize{IRAS Point Source FIR luminosity versus total H$_2$O luminosity of H$_2$O detected galaxies 
(cf. Table 4). For NGC~2146, the total logarithmic H$_2$O luminosity is 0.9 in solar units (Tarchi et al. 2002a). 
Stars denote the ten sources belonging to the FIR selected maser sample. The diagonal line shows the correlation found 
by Jaffe et al. (1981) for individual galactic star forming regions.} 
\label{firh2o}}
\end{figure}

While there is significant scatter (among the sources of Table 2, the most extreme source by far is NGC~3079, whose H$_2$O 
maser would be detectable even at a distance corresponding to $S_{\rm 100\mu m}$$\sim$10\,Jy), we conclude that at present 
sensitivities, there is for most sources a detection threshold near $S_{\rm 100\mu m}$=100\,Jy. It appears that $S_{\rm 100\mu m}$ 
and H$_2$O peak fluxes are roughly proportional, as was already suggested by HWB on the basis of a smaller number of detected 
sources. Such a result is reminiscent of the $L_{\rm FIR}$ -- $L_{\rm H_2O}$ correlation found by Jaffe et al. (1981) for 
galactic star forming regions and is readily explained if most of the detected sources in our FIR sample are associated 
with sites of massive star formation. Four of the ten detected sources are indeed related to star formation, two to AGN, while 
the nature of the remaining four is uncertain. While the scatter is large, nevertheless even the AGN related megamaser galaxies 
roughly follow the correlation found for galactic sources, i.e. $L_{\rm FIR}$/$L_{\rm H_2O}$$\sim$10$^9$ (see Fig.\,\ref{firh2o}). 
This is difficult to explain and might be caused by a spatially extended cascade of nuclear bars that contains warm dust and
that is needed to fuel the very nuclear region (e.g. Shlosman \& Heller 2002).

Given the correlation between FIR and H$_2$O maser flux densities, an improvement in sensitivity by one order of magnitude
to detect H$_2$O masers would lower the 100\,$\mu$m flux threshold from $\sim$100\,Jy to $\sim$10\,Jy, and would provide a 
$\sim$25 times richer sample of detectable targets ($\sim$250 sources at $\delta$$>$--30$^{\circ}$). This enlarged sample 
might then also include some of the brighter ULIRGs that are not part of this study because of too large distances and 
correspondingly low infrared flux densities.

\subsection{The entire extragalactic maser sample}

\subsubsection{The distance bias}

Table 4 lists the 53 galaxies with a total of 57 groups of H$_2$O masers detected beyond the Magellanic Clouds. The 
separation between megamasers (AGN environment) and kilomasers (mainly star formation) is not entirely clear. NGC~2782, 
an $L_{\rm H_2O}$$\sim$12\,L$_{\odot}$ maser, {\it might} not possess an AGN (e.g. Braatz et al. 2004), while the masers 
in NGC~2146 (a total of $\sim$8\,L$_{\odot}$; see Table 5) are known to be related to star formation (Tarchi et al. 2002b). 
A clear separation between jet and accretion disk masers would also be appropriate. The lack of high resolution data toward 
most sources, however, makes such a classification elusive. We thus group the masers according to their isotropic luminosity, 
assuming that all sources with $L_{\rm H_2O}$$\geq$10\,L$_{\odot}$ (the megamasers) are nuclear. 

The H$_2$O detections presented in Table 4 were collected from various surveys with different sensitivities and even 
within a survey, noise levels may differ from source to source. Furthermore, the masers are time variable and it is always 
more difficult to detect a broad weak feature than a stronger but narrower spectral component. In view of this highly 
heterogeneous data base, we adopt a characteristic linewidth of the dominant spectral feature of 20\,km\,s$^{-1}$ and a 
detection threshold of 50\,mJy (this sensitivity is inferior to that in our surveys (Sects.\,4.1 and 4.2) and reflects the 
higher noise levels of most other data, the exception being the Braatz et al. 2004 survey). We should then be able to detect 
masers with 1, 10, 100, 1000, and 10000\,L$_{\odot}$ out to maximal distances of 
$$
D/{\rm Mpc} = [(L_{\rm H_2O}/L_{\odot}) /(0.023 \times\ S/{\rm Jy} \times \Delta 
               V/{\rm km\,s^{-1}})]^{1/2},
$$
i.e. 6.5, 21, 65, 210, and 650\,Mpc, respectively. Note that these distance limits only depend on the square root of the 
adopted observational sensitivity.

\begin{table*}
\label{maser}
\begin{scriptsize}
\begin{center}
\caption{Extragalactic H$_2$O masers beyond the Magellanic Clouds}
\begin{tabular}{lrrrrrrcc}
\hline
Source &  R.A. &  Dec.  & $V_{\rm sys}$           & \multicolumn{1}{c}{$D$}         & log\,$L_{\rm FIR}^{\rm a)}$ & 
                          $T_{\rm dust}^{\rm a)}$ & log\,$L_{\rm H_2O}^{\rm a)}$    & Ref.$^{\rm b)}$ \\
       &  \multicolumn{2}{c}{(J2000)}   & (c$z$) & (Mpc) & (log\,L$_{\odot}$) & (K) & (log\,L$_{\odot}$) \\
\hline
\\
IC\,10                & 00 20 17.9 & +59 18 31 &--350 &  1.2& 8.2& 40&--1.7&   1 \\
                      & 00 20 27.0 & +59 17 29 &      &     &    &   &--0.8& 2,3 \\
NGC\,253              & 00 47 33.1 &--25 17 17 &  240 &  3.0&10.3& 52&--0.8& 4,5 \\
                      & 00 47 33.6 &--25 17 14 &      &     &    &   &--1.7&   5 \\
NGC\,262 (Mrk348)     & 00 48 47.1 & +31 57 25 & 4505 & 62.0&10.6& 42&  2.6& 6,7 \\
IRAS F0106--8034      & 01 07 00.9 &--80 18 24 & 5045 & 67.0&10.7& 32&  2.7&   8 \\
NGC\,449 (Mrk\,1)     & 01 16 07.2 & +33 05 22 & 4780 & 64.0&10.6& 55&  1.7&   9 \\
NGC\,598 (M\,33)      & 01 33 16.5 & +30 52 50 &--180 &  0.7& 9.0& 36&--0.5&10,11\\
                      & 01 33 29.4 & +30 31 55 &      &     &    &   &--1.5&  12 \\
NGC\,591 (Mrk\,1157)  & 01 33 31.2 & +35 40 06 & 4555 & 61.0&10.5& 46&  1.4&  13 \\
NGC\,1052             & 02 41 04.8 &--08 15 21 & 1470 & 17.0& 9.1& 54&  2.1& 9,14\\
NGC\,1068             & 02 42 40.7 &--00 00 48 & 1135 & 14.5&11.2& 54&  2.2&15,16\\ 
Mrk\,1066             & 02 59 58.6 & +36 49 14 & 3600 & 48.0&10.9& 55&  1.5&13,17\\ 
NGC\,1386             & 03 36 46.4 &--36 00 02 &  870 & 17.0& 9.8& 46&  2.1&  18 \\ 
IC\,342$^{\rm c)}$    & 03 46 46.3 & +68 05 46 &   40 &  2.0& 9.0& 49&--2.0&  19 \\ 
UGC\,3255             & 05 09 50.2 & +07 29 00 & 5675 & 75.0&10.5& 40&  1.2&  13 \\
Mrk\,3                & 06 15 36.3 & +71 02 15 & 4010 & 54.0&10.7& 69&  1.0&  13 \\
NGC\,2146             & 06 18 36.6 & +78 21 28 &  900 & 14.5&10.9& 53&  0.0&  20 \\ 
                      & 06 18 38.6 & +78 21 24 &      &     &    &   &  0.0&  20 \\
Mrk\,78               & 07 42 41.7 & +65 10 37 &11195 &150.0&11.0& 60&  1.5&  13 \\
Mrk\,1210             & 08 04 05.8 & +05 06 50 & 4045 & 54.0&10.5& 75&  1.9&   9 \\
NGC\,2639             & 08 43 38.1 & +50 12 20 & 3335 & 44.0&10.4& 34&  1.4& 9,21\\
NGC\,2782             & 09 14 05.1 & +40 06 49 & 2560 & 34.0&10.5& 47&  1.1&  13 \\
NGC\,2824 (Mrk\,394)  & 09 19 02.2 & +26 16 12 & 2760 & 37.0& 9.7& 47&  2.7&  22 \\
NGC\,2960 (Mrk\,1419) & 09 40 36.4 & +03 34 37 & 4930 & 66.0&10.5& 38&  2.6&  23 \\
NGC\,2979             & 09 43 08.5 &--10 23 01 & 2720 & 36.0& 9.9& 38&  2.1&  22 \\ 
NGC\,3034 (M\,82)     & 09 55 52.2 & +69 40 47 &  200 &  3.7&10.6& 65&  0.0&15,24\\
NGC\,3079             & 10 01 57.8 & +55 40 47 & 1120 & 15.5&10.6& 42&  2.7&25,26,27\\
IC\,2560              & 10 16 18.7 &--33 33 50 & 2925 & 35.0&10.2& 47&  2.0&18,28\\
Mrk\,34               & 10 34 08.6 & +60 01 52 &15140 &205.0&11.2& 55&  3.0&  17 \\
NGC\,3556             & 11 11 31.2 & +55 40 25 &  700 & 12.0&10.5& 38&  0.0&  17 \\
Arp\,299 (Mrk\,171)   & 11 28 32.2 & +58 33 44 & 3120 & 42.0&11.7& 61&  2.1&  17 \\
NGC\,3735             & 11 35 57.3 & +70 32 09 & 2695 & 36.0&10.6& 38&  1.3&  29 \\
NGC\,4051             & 12 03 09.6 & +44 31 53 &  730 & 10.0& 9.6& 38&  0.3&  30 \\
NGC\,4151             & 12 10 32.6 & +39 24 21 & 1000 & 13.5&\nodata&\nodata&--0.2&  13 \\ 
NGC\,4258             & 12 18 57.5 & +47 18 14 &  450 &  7.2& 9.9& 33&  1.9&15,31\\
NGC\,4388             & 12 25 46.7 & +12 39 44 & 2520 & 34.0&10.7& 47&  1.1&  13 \\
ESO\,269--G012        & 12 56 40.7 &--46 55 31 & 4950 & 66.0&\nodata&\nodata&  3.0&  22 \\
NGC\,4922             & 13 01 25.2 & +29 18 50 & 7080 & 95.0&11.2& 61&  2.3&  13 \\
NGC\,4945             & 13 05 27.5 &--49 28 06 &  560 &  4.0&10.3& 45&  1.7&32,33\\
NGC\,5194 (M\,51)     & 13 29 52.7 & +47 11 43 &  450 & 10.0&10.3& 33&--0.2& 4,34\\
NGC\,5256 (Mrk\,266)  & 13 38 17.2 & +48 16 32 & 8365 &112.0&11.5& 46&  1.5&  13 \\ 
NGC\,5347             & 13 53 17.8 & +33 29 27 & 2335 & 31.0& 9.9& 44&  1.5&  18 \\
Circinus              & 14 13 09.3 &--65 20 21 &  450 &  4.0&10.1& 54&  1.3&35,36\\
NGC\,5506 (Mrk\,1376) & 14 13 14.8 &--03 12 27 & 1850 & 25.0&10.3& 59&  1.7&   9 \\
NGC\,5643             & 14 32 40.7 &--44 10 28 & 1200 & 16.0&10.3& 39&  1.4&  22 \\
NGC\,5728             & 14 42 23.9 &--17 15 11 & 2795 & 37.0&10.6& 45&  1.9&  13 \\
NGC\,5793             & 14 59 24.7 &--16 41 36 & 3490 & 47.0&10.6& 47&  2.0&37,38\\
NGC\,6240             & 16 52 58.1 & +02 23 50 & 7340 & 98.0&11.8& 58&  1.6&39,40,41,42\\ 
NGC\,6300             & 17 17 00.3 &--62 49 15 & 1110 & 15.0&10.2& 36&  0.5&  22 \\ 
NGC\,6323             & 17 13 18.0 & +43 46 56 & 7790 &104.0&\nodata&\nodata&  2.7&  13 \\
ESO103--G035          & 18 38 20.3 &--65 25 42 & 3985 & 53.0&10.5&121&  2.6&  18 \\
IRAS F19370--0131     & 19 39 38.9 &--01 24 33 & 6000 & 80.0&10.7& 59&  2.2&  22 \\
3C403                 & 19 52 15.8 & +02 30 24 &17690 &235.0&11.2&\nodata&  3.3&  43 \\
NGC\,6926             & 20 31 38.7 &--80 49 58 & 5970 & 80.0&11.1& 39&  2.7&  22 \\
TXS2226--184          & 22 29 12.5 &--18 10 47 & 7495 &100.0&\nodata&\nodata&  3.8&  44 \\
IC\,1481              & 23 19 25.1 & +05 54 21 & 6120 & 82.0&10.4& 65&  2.5&  18 \\
\\
\hline
\end{tabular}
\end{center}
a) For the determination of $L_{\rm FIR}$ and $T_{\rm dust}$ (60/100$\mu$m color 
   temperatures), see Wouterloot \& Walmsley (1986). The IRAS fluxes were taken from Fullmer 
   \& Lonsdale (1989) and, for a few sources (NGC\,598, NGC\,4258, IRAS~F19370--1031 and 3C403), 
   from NED. While, as already noted by HWB and Braatz et al. (1997), the derived dust 
   temperatures are mostly well above 30\,K and thus rather large, a correlation between 
   $T_{\rm dust}$ and $L_{\rm H_2O}$ is not apparent. \\
b) References: (1) Henkel et al. (1986) (2) Becker et al. (1993) (3) Argon et al. (1994)
   (4) Ho et al. (1987) (5) Henkel et al. (2004) (6) Falcke et al. (2000) (7) Peck et al. 
   (2003) (8) Greenhill et al. (2002) (9) Braatz et al. (1994) (10) Churchwell et al. (1977) 
   (11) Greenhill et al. (1993) (12) Huchtmeier et al. (1978) (13) Braatz et al. (2004) (14) 
   Claussen et al. (1998) (15) Claussen et al. (1984) (16) Gallimore et al. (2001) (17) This 
   paper (18) Braatz et al. (1996) (19) Tarchi et al. (2002a) (20) Tarchi et al. (2002b) (21) 
   Wilson et al. (1995) (22) Greenhill et al. (2003b) (23) Henkel et al. (2002) (24) Baudry 
   \& Brouillet (1996) (25) Henkel et al. (1984) (26) Haschick \& Baan (1985) (27) Trotter et 
   al. (1998) (28) Ishihara et al. (2001) (29) Greenhill et al. (1997a) (30) Hagiwara et al.
   (2003b) (31) Herrnstein et al. (1999) (32) Dos Santos \& L{\'e}pine (1979) (33) Greenhill 
   et al. (1997b) (34) Hagiwara et al. (2001b) (35) Gardner \& Whiteoak (1982) (36) Greenhill 
   et al. (2003a) (37) Hagiwara et al. (1997) (38) Hagiwara et al. (2001a) (39) Hagiwara et al. 
   (2002) (40) Nakai et al. (2002) (41) Braatz et al. (2003) (42) Hagiwara et al. (2003a) (43) 
   Tarchi et al. (2003) (44) Koekemoer et al. (1995) \\ 
c) The maser luminosity refers to a brief flaring episode \\
\end{scriptsize}
\end{table*}

We can check how consistent this is with the sample of detected masers listed in Table 4. The total H$_2$O 
luminosity per galaxy is taken. Table 5 shows the results. The number of detections in the most likely distance bin 
is given in italics. IC~342 was not included because of the intrinsic weakness of its maser. It is apparent that for 
the kilomaser galaxies (here defined to show luminosities $<$10\,L$_{\odot}$) either the number of sources in the 
expected bin is by far the highest or there are additional detections at both lower and higher distances (the latter 
a consequence of the fact that the sensitivity of the surveys is not uniform). This provides a picture that is 
approximately consistent with expectations.

For the more luminous megamasers the situation is different. The distance distribution of the 20 masers with 10\,L$_{\odot}$ 
$\leq$ $L_{\rm H_2O}$ $<$ 100\,L$_{\odot}$ is still consistent. Four are located at $D$$<$21\,Mpc, and four at a distance 
higher than the estimated limiting distance of 65\,Mpc. Most of the detections are obtained in their most likely distance bin. 
Among the 18 masers with 100\,L$_{\odot}$ $\leq$ $L_{\rm H_2O}$ $<$ 1000\,L$_{\odot}$, however, 13 are closer or near
$D$$\sim$65\,Mpc, the inner limit of the most likely distance range, while among the four galaxies with 1000\,L$_{\odot}$ 
$\leq$ $L_{\rm H_2O}$ $<$ 10000\,L$_{\odot}$, three are closer than the corresponding $D$$\sim$210\,Mpc limit. The fourth 
megamaser, 3C\,403, surpasses this limit by only a small amount. None of the masers in the last two groups has a distance 
larger than the estimated maximum distance.

The surveys do not cover the entire sky and are therefore incomplete by an unknown amount. The fact that distances of 
masers with lower luminosity are consistent with expected values indicates that a bias related to the distance of these 
sources is negligible. Most of the luminous megamasers ($L_{\rm H_2O}$$\geq$100\,L$_{\odot}$), however, are observed at 
distances that are smaller than expected. Is this an effect of different lineshapes or a consequence of the observed 
sample of sources? Considering the four most luminous targets with $L_{\rm H_2O}$ $\geq$ 1000\,L$_{\odot}$, only 
TXS2226--184 has an unusually wide profile, while the others show `normal' lineshapes. Without going into any detail, 
we note that the situation is similar for sources with 100\,L$_{\odot}$ $\leq$ $L_{\rm H_2O}$ $<$ 1000\,L$_{\odot}$. 
We thus conclude that the bias towards `nearby' sources in the sample of luminous water masers is caused by the 
selection of galaxies so far observed. The entire megamaser sample is dominated by the surveys of Braatz et al. (1994, 
1996, 1997, 2003, 2004) that are mostly confined to recessional velocites $\la$7000\,km\,s$^{-1}$, i.e. out to 
$D$$\sim$100\,Mpc (this also holds for the two surveys discussed in Sects.\,4.1 and 4.2). Sources with significantly 
larger distances were rarely observed.

From the number of sources at `near' distances we may extrapolate to the larger volumes to estimate the percentage of 
missing detections in this larger volume. This may provide lower limits because detections at the `nearby' distances may 
be incomplete as well. For the more luminous megamasers with 100\,L$_{\odot}$ $\leq$ $L_{\rm H_2O}$ $<$ 1000\,L$_{\odot}$, 
four sources are observed within $D$=21\,Mpc, so that $\sim$4000$\pm$2000 detectable targets may be expected within 
$D$=210\,Mpc. This has to be compared with 16 known such objects. With 11 known sources within $\sim$65\,Mpc, we still 
expect $\sim$350$\pm$105 objects within $D$=210\,Mpc, a factor of $\sim$20 above the detected number. Among the most 
luminous four sources, those with $L_{\rm H_2O}$ $\ga$ 1000\,L$_{\odot}$, three are detected inside of 210\,Mpc, so we 
would expect 100$\pm$60 detectable targets out to $D$=650\,Mpc, of which so far only four have been identified.

\begin{table}
\label{Detections}
\begin{scriptsize}
\begin{center}
\caption{Number of detected maser galaxies per luminosity and distance interval (see Table 4)$^{\rm a)}$}
\begin{tabular}{ccccccc}
\hline
\\
log\,($L_{\rm H_2O}$) & \multicolumn{6}{c}{Distance}                                \\
(L$_{\odot}$)         & \multicolumn{6}{c}{(Mpc)}                                   \\
\\
\hline
                     & $<$2   &   2--6   &  6--21    & 21--65 & 65--210 & 210--650  \\
\hline
\\
--1.0-(--0.1)        &    2   & {\it 1}  &     2     &   --   &   --    &    --     \\
0.0--0.9             &   --   &    1     &  {\it 4}  &   --   &   --    &    --     \\
1.0--1.9             &   --   &    2     &     2     &{\it 12}&    4    &    --     \\
2.0--2.9             &   --   &   --     &     4     &    7   & {\it 7} &    --     \\
3.0--3.9             &   --   &   --     &    --     &   --   &    3    & {\it 1}   \\
\\
\hline
\end{tabular}
\end{center}
a) In italics: Expected highest distance bin where sources of a given luminosity should still be detectable. Because this 
distance bin contains a much larger volume than the nearer ones, most masers are expected to be there. For each of the galaxies 
in Table 4 the total integrated 22\,GHz H$_2$O luminosity was taken. This amounts to an integrated single-dish flux density 
corresponding to $\sim$8\,L$_{\odot}$ in the case of NGC~2146, where VLA data from Tarchi et al. (2002b) revealed two 
$\sim$1\,L$_{\odot}$ masers (see Table 4).\\
\end{scriptsize}
\end{table}

While the large errors in the predicted numbers of detectable sources may raise scepticism, an analysis of the distances 
of the galaxies belonging to the two most luminous H$_2$O luminosity bins (100\,L$_{\odot}$ $\leq$ $L_{\rm H_2O}$ $<$ 
10000\,L$_{\odot}$) yields a clear result. 14 of the 22 sources in these bins are not in the expected most distant shell 
(between $D$=65 and 210 or between $D$=210 and 650\,Mpc, respectively) but are located more nearby. Two additional sources 
are located at the inner boundary of the most likely shell, while no source is detected beyond the estimated distance limit. 
This implies that the majority of sources, 73\%, is located within a volume that encompasses only 3.2\% of the volume in which 
the masers would be detectable. Assuming an isotropic spatial distribution and applying the Bernoulli theorem, a deviation of 
8\% from 73\% corresponds to 1$\sigma$. The discrepancy between the expected (3.2\% of the detections in the inner, 96.8\% 
in the outer shell) and observed (73\% in the inner, 27\% in the outer shell) spatial distributions is therefore significant. 
We conclude that statistical evidence strongly indicates that only {\it a tiny fraction of the luminous megamaser sources 
detectable with presently available instrumentation has been discovered to date}. 

So far we have not yet considered that the maser luminosities are not necessarily at the upper edge of their respective bin. 
This has the consequence that not all of them should be detectable out to the upper limit of the corresponding most likely 
distance interval. To quantify this we have to determine the H$_2$O luminosity function.

\subsubsection{The H$_2$O maser luminosity function}

The luminosity function $\Phi(L_{\rm H_2O})$ is the number density of objects with luminosity $L_{\rm H_2O}$ per logarithmic 
interval in $L_{\rm H_2O}$. An unbiased direct measurement of $\Phi(L_{\rm H_2O})$ would require that all objects with a given 
luminosity be detected within the survey volume, which is not possible in flux limited surveys like those presented in 
Sects.\,4.1 and 4.2. Instead, each object in a survey has an effective volume in which it could have been detected and the 
sum of detections weighted by their available volumes $V_{\rm i}$ determines the luminosity function. 

Not accounting for the incompleteness of the detected H$_2$O megamaser sample and ignoring the possibility that in different 
luminosity bins the fraction of detected sources may be different, we can derive a zeroth order approximation to the luminosity 
function (LF) of extragalactic H$_2$O maser sources. Such a computation is not only limited by the effects mentioned above but 
there are additional factors, the main three being: 
\newline
(a) Non-uniform sky coverage resulting from several large and many small surveys. Most of the large surveys have used 
    optical-magnitude-limited samples of galaxies in the northern sky. We have approximated the sky coverage to be the entire 
    northern sky; 
\newline 
(b) Different sensitivity limits in the various surveys. We have considered two typical detection thresholds: 1\,Jy\,km\,s$^{-1}$ 
    (from e.g. a detection limit of a 20\,km\,s$^{-1}$ line with peak flux 50\,mJy) and 0.2\,Jy\,km\,s$^{-1}$ (i.e. five times 
    weaker).
\newline
(c) The use of optical-magnitude limited galaxy samples for H$_2$O maser surveys. Given this, the H$_2$O maser LF is best 
    calculated via the optical LF and the bivariate H$_2$O and optical LF (e.g. Meurs \& Wilson 1984). Given the diversity of 
    the selection criteria in different surveys, however, we are forced to directly calculate the H$_2$O LF.

We use the standard $V/V_{\rm max}$ method (Schmidt 1968) to estimate the H$_2$O LF. The sample objects (Table~4) were divided 
into bins of 0.5\,dex over the range $L_{\rm H_2O}$ = 10$^{-1}$--10$^{4}$\,L$_{\odot}$. For each luminosity bin ($L_{\rm p}$; 
$p$=1,10) we calculated the differential LF value as follows:
$$
\Phi(L_{\rm p}) = \frac{4\pi}{\Omega} \Sigma_{\rm i=1}^{n(L_{\rm p)}} (1/V_{\rm max})_{\rm i}
$$

\begin{figure}[ht]
\hspace{-0.6cm}
\includegraphics[angle=0,width=9.5cm]{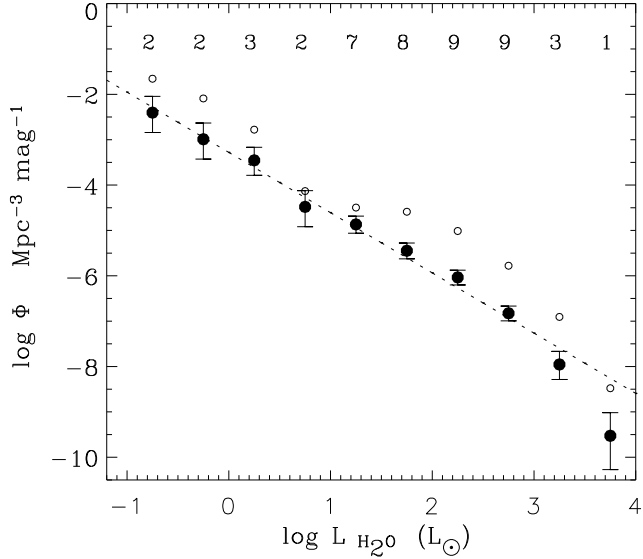}
\vspace{-0.4cm}
\caption[]{\footnotesize{The 22\,GHz H$_2$O luminosity function (LF) for maser galaxies beyond the Magellanic Clouds (Table 4). 
The filled circles show the LF for a detection limit of 0.2\,Jy\,km\,s$^{-1}$. The dashed line shows a fit with a slope
of --1.3 (for uncertainties in the slope, see Sect.\,4.3.2). Error bars of individual points are derived from Poisson statistics 
following Condon (1989). Not included in the diagram are IC~342 (too low maser luminosity) and UGC~3255, Mrk~3, Mrk~78, 
NGC~4151, NGC~5256 and NGC~6240 (below the adopted detection limit). The number of masing galaxies in each bin are also shown. 
To illustrate the effect of changing the sensitivity, we also show the LF for a detection limit of 1\,Jy\,km\,s$^{-1}$ (empty 
circles; in this case 19 of the 53 galaxies fall below the detection limit).}
\label{maserlf}}
\end{figure}

Here $n(L_{\rm p})$ is the number of galaxies with $L_{\rm p}$--0.25 $<$ log\,$L_{\rm H_2O}$ $<$ $L_{\rm p}$+0.25. The term after 
the summation sign represents the inverse of the maximum volume ($V_{\rm max}$) over which an individual galaxy can be detected 
given its maser luminosity and the detection limit of the survey. As discussed above in (b), we consider two different detection 
limits: 0.2\,Jy\,km\,s$^{-1}$ and 1\,Jy\,\kms. In the former case, 6 masers have to be left out of the computation since they are 
weaker than the assumed detection limit, and in the latter case 19 masers have to be omitted. The maser in IC~342 is outside the 
considered H$_2$O luminosity range. Results are plotted in Fig.~\ref{maserlf} for both detection limits.

Fig.\,\ref{maserlf} demonstrates that the H$_2$O luminosity function does not strongly depend on the detection limit used. From 
the overall slope of the luminosity function we derive $\Phi$ $\propto$ (log\,$L_{\rm H_2O})^{-1.6}$ which is steeper than 
the LF for OH megamasers (see Darling \& Giovanelli 2002b). Noteworthy is the fast decay in the number of sources at the upper end 
of the maser luminosity function that could indicate that ultraluminous H$_2$O `gigamasers' are rare. Obvious is also a low number 
of sources in the 1--10\,L$_{\odot}$ bin. This bin marks the upper end of the luminosity distribution of known star forming regions 
and is located slightly below the luminosity of the `weak' megamaser sources. So there might exist a minimum of H$_2$O emitting 
targets just below the megamaser luminosity threshold. Both results, the minimum at $L_{\rm H_2O}$ = 1--10\,L$_{\odot}$ and the 
fast decline at highest maser luminosities, are, however, of questionable significance. The number of sources in the $L_{\rm H_2O}$ 
= 0.1--10\,L$_{\odot}$ bins is not yet large enough to make a convincing case. And Fig.\,\ref{maserlf} does not account for the 
distance bias discussed for the most luminous sources in Sect.\,4.3.1.

Can we hope to see H$_2$O megamaser emission at cosmological distances? The most luminous megamaser known at present, that in 
TXS2226--184 with a redshift of $z$=0.025 (Koekemoer et al. 1995), a peak flux density of 400\,mJy, and a linewidth of 
$\sim$80\,km\,s$^{-1}$ would be detectable by a 100-m telescope out to $z$$\sim$0.4. With the Square Kilometer Array (SKA)
detections at significant redshifts will thus be possible.  

We may also estimate the number of detectable H$_2$O megamasers with the LF shown in Fig.\,\ref{maserlf}. The lower limit to the 
H$_2$O luminosity of a source at distance $D$ varies as
$$
L_{\rm H_2O} =  {\rm C_1} D_{\rm max}^2.
$$
The total number of H$_2$O masers of a given luminosity to be detected within this distance can then be expressed by 
$$
N_{\rm tot, H_2O} = \int{N_{\rm H_2O}\ {\rm d}V_{\rm obs}} = (4/3) \pi \,N_{\rm H_2O}\,D_{\rm max}^3, 
$$
where $N_{\rm H_2O}$ is the uniform space density of the masers. Assuming that the overall slope of the LF is independent of the 
maser luminosity, we then obtain with
$$
N_{\rm H_2O} = {\rm C_2} L_{\rm H_2O}^{-1.6}
$$
(see above)
$$
N_{\rm tot, H_2O} = (4/3)\pi \,{\rm C_1}^{-1.5}\,{\rm C_2}\,\,L_{\rm H_2O}^{-0.1} \propto\ L_{\rm H_2O}^{-0.1}
$$
This implies that the number of observable masers is almost independent of their intrinsic luminosity: The smaller source density 
at higher H$_2$O luminosities is compensated by the larger volume in which they can be detected. In principle, this would
permit H$_2$O detections with 100-m sized telescopes out to large redshifts, provided that the LF is not steepening at very high 
maser luminosities and that it is possible to find suitable candidate sources.
 
In Sect.\,4.3.1 evidence was found that only a small fraction of the detectable luminous H$_2$O maser sources is known to date. 
This may be the main cause for the comparatively steep slope in the LF at highest maser luminosities (see Fig.\,\ref{maserlf}). 
Ignoring therefore the two bins with highest maser luminosities in Fig.\,\ref{maserlf}, the slope of the LF becomes --1.3 instead 
of --1.6. This provides a realistic estimate of systematic errors but does not qualitatively change our conclusion. 

At the end of Sect.\,4.3.1 it was mentioned that the LF is also needed to quantify the deficit of distant high luminosity masers.
Adopting the result (see above) that the number of detectable masers is almost independent of the maser luminosity, we can 
determine with 
$$
V/V_{\rm i+1} = [L_{\rm H_2O,i+1}-L_{\rm H_2O,i}]^{-1} L_{\rm H_2O,i+1}^{-1.5} \int{L_{\rm H_2O}^{1.5}\,\,{\rm d}L_{\rm H_2O}}
$$ 
the fractional volume $V_{\rm f}$ = $V$/$V_{\rm i+1}$ that masers can occupy in the luminosity range $L_{\rm H_2O,i}$ to 
$L_{\rm H_2O,i+1}$ relative to the maximum volume $V_{\rm i+1}$ defined by the upper luminosity limit $L_{\rm H_2O,i+1}$. 
The indices $i$ and $i$+1 indicate the lower and upper boundaries of the studied luminosity bin. The integral is calculated 
between the limits $L_{\rm H_2O,i}$ and $L_{\rm H_2O,i+1}$. For order of magnitude bins as considered in Sect.\,4.3.1 an 
occupied average volume of 44\% is reached. For the luminous megamasers with $L_{\rm H_2O}$$\geq$100\,L$_{\odot}$ we thus 
expect 7\% of the detections in the inner shells and 93\% in the outer envelope that is still far from the 
observed (73\% versus 27\%) maser distribution. We thus conclude that the distance bias outlined in Sect.\,4.3.1 is real 
and that the LF slope remains approximately constant within the $L_{\rm H_2O}$ range shown in Fig.\,\ref{maserlf}.

\section{Conclusions}

This article presents a search for 22\,GHz ($\lambda$$\sim$1.3\,cm) H$_2$O masers towards two classes of objects, i.e. 
galaxies that (1) either contain nuclear jets that are oriented close to the disk of the galaxy and the plane of the sky
or that (2) are bright in the far infrared. The main results are:

(1) Two new `jet-maser' sources were detected. One of these, Mrk\,1066, shows two components that bracket the systemic 
velocity of its parent galaxy. Mrk~34 contains the most distant and one of the most luminous megamasers so far observed in 
a Seyfert galaxy. The source comprises three spectral components that cover a velocity range of $\sim$1000\,km\,s$^{-1}$. 

(2) Two new masers were also detected in the sample of FIR bright galaxies. One source is a relatively weak ($L_{\rm H_2O}$ 
$\sim$ 1\,L$_{\odot}$) kilomaser, while the other, Arp~299, is a luminous megamaser in a merging system with high infrared 
luminosity. There may be two maser components, one associated with the subsystem IC~694 and the other with NGC~3690,
following the conventional nomenclature of the source.

(3) When compared with other H$_2$O surveys, the jet-maser and FIR maser samples show extremely high detection rates and are 
thus providing a strong motivation for further studies. Including previously detected sources, the jet-maser detection rate 
is 50\% (7/14), while the FIR maser detection rate is 22\% (10/45). 

(4) As far as one can judge from single-dish data (this paper and Braatz et al. 2004), a significant fraction of the `jet-maser' 
sources appear to be disk-masers with a closer resemblance to NGC~4258 than to classical jet-maser sources like Mrk~348 or NGC~1052. 

(5) The detection rate in the sample of bright FIR sources is a function of the FIR flux density. This implies that more 
sensitive surveys will detect H$_2$O in galaxies with smaller 100$\mu$m fluxes. An increase in observational sensitivity by 
a factor of $\sim$10 should yield a 25-fold increase in the number of detections.

(6) The correlation between IRAS Point Source and total H$_2$O luminosity of a galaxy follows, with significant scatter,
the correlation found for individual star forming regions in the Galaxy. While the agreement is expected for galaxies hosting
masers associated with star formation, the agreement for galaxies with AGN dominated masers is less obvious. It may be related
to spatially extended dust rich cascades of bars that fuel the central engine.

(7) 60$\mu$m/100$\mu$m color temperatures from the IRAS Point Source Catalog are not correlated with H$_2$O maser luminosities
(see footnote a) in Table 4). 

(8) There is an observational distance bias: Most of the detectable luminous H$_2$O megamasers ($L_{\rm H_2O}$$>$100\,L$_{\odot}$) 
have not yet been found.

(9) The extragalactic H$_2$O maser luminosity function (LF) might show a minimum near the transition between the luminosity
range of kilomasers (mostly star formation) and megamasers (AGN) in the interval 1--10\,L$_{\odot}$. The overall slope is 
$\sim$--1.6 and implies that the number of observable masers is almost independent of their luminosity. If the LF is not steepening 
at very high luminosities and if there is a chance to find suitable candidate sources, masers should be detectable with existing 
telescopes out to cosmological distances.

\begin{acknowledgements}
We wish to thank M. Elitzur and L.J. Greenhill for useful discussions during the conception of this project and an
anonymous referee for carefully reading the draft and making useful suggestions. AP and AT wish to thank the MPIfR for their 
hospitality during the observing run. NN was partially supported by the Italian Ministry for University and Research (MURST) 
under grant Cofin00-02-36 and the Italian Space Agency (ASI) under grant 1/R/27/00. This research has made use of the NASA/IPAC 
Extragalactic Database (NED) which is operated by the Jet Propulsion Laboratory, Caltech, under contract with NASA. This research 
has also made use of NASA's Astrophysics Data System Abstract Service.
\end{acknowledgements}

\clearpage

\end{document}